\documentclass{article}
\usepackage{arxiv}
\renewcommand{\undertitle}{}
\renewcommand{\headeright}{}
\date{}

\usepackage[numbers,sort&compress]{natbib}
\usepackage{caption}

\usepackage{microtype}
\usepackage[normalem]{ulem}
\usepackage{xspace}
\usepackage{xcolor}
\usepackage{tikz}
\usetikzlibrary{pgfplots.groupplots}

\usepackage{pifont}
\usepackage{booktabs}

\usepackage{pgfplots}
\usepackage{pgfplotstable}
\usepackage{amsmath,amsfonts,amsthm,enumitem}
\usepackage{algpseudocode}
\usepackage[linesnumbered,ruled]{algorithm2e}
\usepackage{graphicx}
\usepackage{url}

\usepackage{textcomp}
\usepackage{iftex}
\ifPDFTeX
    \usepackage[T1]{fontenc}
    \usepackage[utf8]{inputenc}
\fi

\usepackage{colortbl}
\usepackage{tcolorbox}
\usepackage{listings}
\usepackage{makecell}

\usetikzlibrary{patterns,arrows,calc}
\usepgfplotslibrary{fillbetween}
\usepackage{multicol}
\usepackage{multirow}
\usepackage{wrapfig}
\usepackage{placeins}
\usetikzlibrary{trees}

\definecolor{toolbg}{gray}{0.92}
\newcommand{\zero}{\textcolor{black!30}{0}}

\pgfplotsset{compat=1.18}
\definecolor{propBlue}{HTML}{3A6EA5}
\definecolor{agentOrange}{HTML}{E8A33D}

\usepackage[]{mdframed}

\definecolor{dkgreen}{rgb}{0,0.6,0}
\definecolor{gray}{rgb}{0.5,0.5,0.5}
\definecolor{mauve}{rgb}{0.58,0,0.82}
\definecolor{orangep}{rgb}{0.71, 0.43, 0.89}
\definecolor{orp}{rgb}{1, 0.7, 0.278}

\definecolor{darkBlue}{rgb}{0.000000,0.000000,0.545098}
\definecolor{darkGreen}{rgb}{0.000000,0.392157,0.000000}
\definecolor{DarkGray}{gray}{0.4}
\definecolor{javared}{rgb}{0.6,0,0}
\definecolor{javagreen}{rgb}{0.25,0.5,0.35}
\definecolor{javapurple}{rgb}{0.5,0,0.35}
\definecolor{javadocblue}{rgb}{0.25,0.35,0.75}
\definecolor{lightgray}{gray}{0.95}
\definecolor{shadecolor}{RGB}{150,150,150}
\definecolor{blueA}{RGB}{204,229,255}
\definecolor{redA}{RGB}{112,0, 0}
\lstdefinestyle{MyCSmallStyle} {
  language=C++,
  frame=none,
  xleftmargin=15pt,
  stepnumber=1,
  numbers=left,
  numbersep=5pt,
  numberstyle=\tiny\color[black]{0.177},
  belowcaptionskip=\bigskipamount,
  captionpos=b,
  escapeinside={*'}{'*},
  tabsize=5,
  emphstyle={\bf},
  escapechar=!,
  basicstyle=\scriptsize\ttfamily,
  keywordstyle=\color{javapurple}\bfseries,
  stringstyle=\color{javared},
  commentstyle=\color{javagreen},
  morecomment=[s][\color{javadocblue}]{/**}{*/},
  showspaces=false,
  columns=flexible,
  showstringspaces=false,
  morecomment=[l]{//},
  tabsize=2,
  breaklines=true,
  moredelim=[is][\underbar]{^}{^}
}
\lstdefinestyle{MyJavaSmallStyle} {
  language=Java,
  frame=none,
  xleftmargin=15pt,
  stepnumber=1,
  numbers=left,
  numbersep=5pt,
  numberstyle=\color{DarkGray},
  belowcaptionskip=\bigskipamount,
  captionpos=b,
  escapeinside={*'}{'*},
  tabsize=5,
  emphstyle={\bf},
  basicstyle=\scriptsize\ttfamily,
  keywordstyle=\color{javapurple}\bfseries,
  stringstyle=\color{javared},
  commentstyle=\color{javagreen},
  morecomment=[s][\color{javadocblue}]{/**}{*/},
  showspaces=false,
  columns=flexible,
  showstringspaces=false,
  morecomment=[l]{//},
  tabsize=2,
  breaklines=true,
  moredelim=[is][\underbar]{^}{^}
}

\lstdefinelanguage{Scala}{
  keywords={typeof, new, true, false, catch,def,val, function, return, null, catch, switch, var, if, in, while, do, else, case, break, assert, static, void ,declare, const, for, define,fun, ite,class, not, check,sat,String, Int, ArrayList},
  keywordstyle=\color{blue}\bfseries,
  ndkeywords={ export,extends, boolean, throw, implements, import, this, abstract,reduceByKey, reduce, filter, map, reduceByKey, join, Join1, public },
  ndkeywordstyle=\color{mauve}\bfseries,
  otherkeywords={+, =>,<=, ==, >,< , ||},
  identifierstyle=\color{black},
  sensitive=false,
  comment=[l]{//},
  morecomment=[s]{/*}{*/},
  commentstyle=\color{purple}\ttfamily,
  stringstyle=\color{red}\ttfamily,
  morestring=[b]',
  morestring=[b]"
}

\definecolor{assertred}{RGB}{180,30,30}

\lstdefinestyle{pythonbase}{
  language=Python,
  basicstyle=\ttfamily\small,
  columns=fullflexible,
  keepspaces=true,
  showstringspaces=false,
  frame=single,
  breaklines=true,
  linewidth=\linewidth,
  keywordstyle={},
}

\lstdefinestyle{pythonassert}{
  style=pythonbase,
  moredelim=**[is][\color{assertred}\bfseries]{@}{@},
}

\newcommand{\MyPara}[1]{\par\vspace{.1em}\noindent\textbf{\textit{#1}}~}
\newcommand{\codefont}[1]{{\texttt{#1}}}

\newcommand{\ie}{\emph{i.e.,}\xspace}

\DeclareRobustCommand{\approachfontsize}{\small}
\newcommand{\toolname}{Propilot}
\newcommand{\agentname}{Agent-Only}
\newcommand{\tool}{{\approachfontsize{\textsc{\toolname}}}\xspace}

\newcommand{\agentonly}{{\approachfontsize{\textsc{\agentname}}}\xspace}
\newcommand{\propopus}{{\approachfontsize{\textsc{\toolname-Opus-5.5}}}\xspace}
\newcommand{\propgptfive}{{\approachfontsize{\textsc{\toolname-GPT-5.5}}}\xspace}
\newcommand{\propgptsix}{{\approachfontsize{\textsc{\toolname-GPT-6.1}}}\xspace}
\newcommand{\agentopus}{{\approachfontsize{\textsc{\agentname-Opus-5.5}}}\xspace}
\newcommand{\agentgptfive}{{\approachfontsize{\textsc{\agentname-GPT-5.5}}}\xspace}
\newcommand{\agentgptsix}{{\approachfontsize{\textsc{\agentname-GPT-6.1}}}\xspace}
\newcommand{\propqwen}{{\approachfontsize{\textsc{\toolname-Qwen-3.8}}}\xspace}
\newcommand{\propdeepseek}{{\approachfontsize{\textsc{\toolname-DeepSeek-4.1}}}\xspace}
\newcommand{\propminimax}{{\approachfontsize{\textsc{\toolname-MiniMax-3.1}}}\xspace}
\newcommand{\agentqwen}{{\approachfontsize{\textsc{\agentname-Qwen-3.8}}}\xspace}
\newcommand{\agentdeepseek}{{\approachfontsize{\textsc{\agentname-DeepSeek-4.1}}}\xspace}
\newcommand{\agentminimax}{{\approachfontsize{\textsc{\agentname-MiniMax-3.1}}}\xspace}
\newcommand{\plusskeleton}{{\approachfontsize{\textsc{\agentname+Skeleton}}}\xspace}
\newcommand{\plusvalidation}{{\approachfontsize{\textsc{\agentname+Skeleton+Validation}}}\xspace}

\AddToHook{cmd/@makecaption/before}{%
    \begingroup
    \let\approachfontsize\relax
}
\AddToHook{cmd/@makecaption/after}{\endgroup}

\newcommand{\rqsummarybox}[2]{%
    \begin{tikzpicture}
        \node[anchor=text,text width=\columnwidth-0.5cm, draw, rounded corners, line width=0.5pt, fill=black!5!white, inner sep=2mm] (big) {\\
            #2
        };
        \node[draw, rounded corners, line width=.5pt, fill=black!30!white, anchor=west, xshift=5mm] (small) at (big.north west) {\bfseries Summary #1};
    \end{tikzpicture}%
}

\NewDocumentEnvironment{rqsummary}{m +b}{
    \par\addvspace{\intextsep}
    \noindent\begin{minipage}{\columnwidth}
        \centering
        \rqsummarybox{#1}{#2}
    \end{minipage}\par
    \addvspace{\intextsep}
}{}

\usepackage[hidelinks]{hyperref}

\def\BibTeX{{\rm B\kern-.05em{\sc i\kern-.025em b}\kern-.08em
    T\kern-.1667em\lower.7ex\hbox{E}\kern-.125emX}}

\title{Tensor-Algebraic Property Skeletons: Guiding Agents to Construct Semantic Oracles for DL Compilers}

\author{
\textbf{
Yuxin Qiu\thanks{Corresponding email: \href{mailto:yuxin.qiu@email.ucr.edu}{yuxin.qiu@email.ucr.edu}.}\textsuperscript{1},
Geyang Xu\textsuperscript{1},
Ben Limpanukorn\textsuperscript{2},
Seongmin Lee\textsuperscript{2},
}\\
\textbf{
Jiyuan Wang\textsuperscript{3},
Miryung Kim\textsuperscript{2},
Qian Zhang\textsuperscript{1}
}\\
{
\textsuperscript{1}UC Riverside
\quad
\textsuperscript{2}UCLA
\quad
\textsuperscript{3}Tulane University
}
}

\begin{document}

\maketitle

\begin{abstract}
Deep learning (DL) compilers optimize tensor computation graphs through transformations that can silently change what a model computes.
Existing work generates well-formed computation graphs and tensor inputs but typically judges results through crashes or differential comparison.
Tensor algebra offers stronger oracles, such as commutativity and reduction decomposition, but turning an identity into an executable check requires selecting applicable operators, valid inputs, and precision-aware comparisons.

We present \tool, an agentic testing framework that guides LLM agents to construct semantic oracles from tensor algebra.
Its core abstraction, the {\em property skeleton}, couples a relation with applicability conditions and an oracle template whose compiler-specific slots the agent fills.
\tool comes with sixteen skeletons and twenty data generators.
It validates each test before execution and uses rejection reasons, execution results, and usage records to steer generation toward untested combinations.

We evaluate \tool with six agents on three DL compilers against \agentonly, which uses the same agents without \tool.
\tool detects 96 distinct compiler root causes, including 25 with silent wrong outputs that \agentonly misses.
It also generates 85.34\% fewer invalid tests and covers at least 2.73$\times$ as many operator--property pairs on each compiler.
On 161 faulty TVM versions, \tool detects 100, against 53 for \agentonly.
\end{abstract}

\section{Introduction}
\label{sec:introdction}
Deep learning (DL) compilers, such as TVM~\cite{tvm} and ONNX-MLIR~\cite{onnx-mlir}, are now part of the critical deployment path for trained models.
They lower tensor computation graphs, which contain tensor operators such as \codefont{matmul} and \codefont{add}, into executable code through compiler transformations such as graph rewrites, operator fusion, and memory layout optimization.
These transformations are necessary for performance, but they also create a correctness risk because an optimized program may compile and run while computing values that no longer match the expected tensor computation~\cite{shen2021study,nnsmith,polyjuice}.

\begin{figure}[t]
    \centering
    \includegraphics[width=\linewidth]{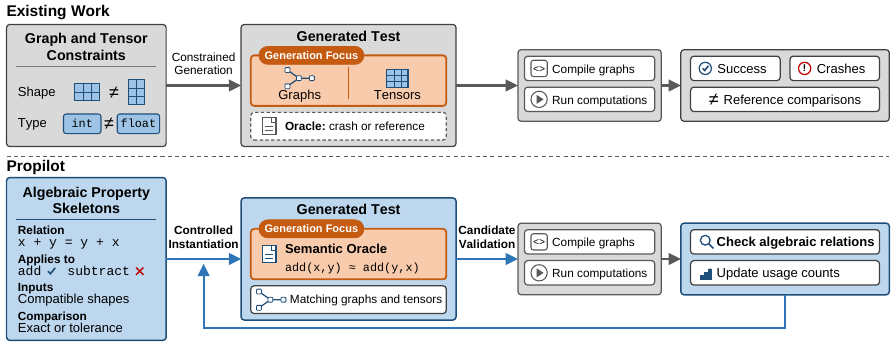}
    \caption{Overview of \tool compared with existing work.
    Existing work generates well-formed computation graphs and tensor inputs from graph and tensor constraints, and it judges the compiled results through crashes or comparison with reference outputs.
    \tool instead starts from the property skeleton.
    It instantiates a property skeleton into a semantic oracle together with the computation graph and tensor inputs on which the oracle must hold, validates each candidate before execution, and feeds the outcomes and updated usage records back to guide the next instantiation.}
    \label{fig:workflow}
\end{figure}

Testing a DL compiler requires two things: inputs that reach its transformations, and an oracle that judges what the compiled program computes.
Existing fuzzing work has made substantial progress on the first by mining, inferring, or solving tensor computation graph constraints, so that generated graphs use supported operators, legal data flows, and valid attributes, and tensor inputs match their shapes, data types, and value domains~\cite{nnsmith,neuri,gencog,deeprel,hirgen,modelmeta,synthfuzz,xcheck}.
The oracles, however, have remained comparatively weak.
Crash-based oracles flag failures during compilation or execution~\cite{tzer,hirgen}, but they cannot see silent semantic drifts, which change model outputs without triggering frontend validation errors or compilation failures~\cite{shen2021study,polyjuice,hirgen}.
Differential oracles compare compiled outputs with a reference implementation or another backend~\cite{nnsmith,neuri,hirgen}, but they require a reference that exists for the operator and agrees with the compiler on its intended semantics.

Tensor algebra offers stronger oracles~\cite{taco-oopsla17,tensor-decomposition-textbook}.
Identities such as commutativity, idempotence, and reduction decomposition state relations that correct outputs must satisfy, so a {\em semantic oracle} built on them can judge a compiled program without an expected output for every tensor input.
Property-based testing (PBT)~\cite{quickcheck,property-based-testing,JQF,Zest} can exercise such relations over generated inputs, but a mathematical relation alone does not specify an executable check.
The tested operators must satisfy the relation's assumptions, the tensor inputs must make it hold, and the assertion must account for numerical precision.
For example, associativity applies to addition, but regrouping floating-point additions can change their rounded results, so an exact-equality check can report a false alarm even when the compiler is correct.
These choices must also remain consistent as the relation is instantiated for different operators and compilers.

Recent work has shown that large language models (LLMs) can generate test programs for complex software systems, including DL libraries and compilers~\cite{titanfuzz,whitefox}, and can write property-based tests from documentation~\cite{vikram2024largelanguagemodelswrite,maaz2025agenticpropertybasedtesting}.
This capability could automate the compiler-specific parts of semantic oracles, but a generated test can compile and run even when its assertion uses an inapplicable property or an incorrect numerical tolerance.
Property templates~\cite{lee2026propertytemplates} reduce this risk in data-intensive computing by fixing a test harness and workload generators and leaving only property-specific slots to the agent.
DL compilers, however, differ in their input formats and operator sets, so the agent must also choose the operator, construct the tensor inputs, and adapt the oracle to each compiler's interface.
The central challenge is therefore to guide an agent to construct executable semantic oracles from tensor algebra together with the computation graphs and tensor inputs needed to exercise them.

\MyPara{\textup{\tool.}}
In this paper, we propose \tool, an LLM-driven agentic testing framework that guides agents to construct semantic oracles from tensor algebra, such that the generated tests check algebraic properties across the compiler's operators and reveal wrong outputs that compile and run without error.
Its core insight is to record, once, the conditions that make an oracle valid together with the relation it checks and to reuse them for every generated test.
While existing work emphasizes generating well-formed computation graphs and tensor inputs, \tool focuses on constructing the oracles that check the compiled results, as shown in Figure~\ref{fig:workflow}.
\tool realizes this idea through {\em property skeletons}, each of which couples a semantic relation with operator applicability conditions, tensor input constraints, and an oracle template whose compiler-specific slots the agent fills.
Its 16 skeletons center on algebraic and tensor properties, including three fallback skeletons, such as comparison with a reference implementation, to extend testing to operators that no algebraic or tensor relation covers.
These specifications constrain the agent's choice of operators, tensor inputs, and output comparisons, while the agent adapts the test code to the target compiler.
\tool's workflow has three stages.

First, \tool performs controlled instantiation of property skeletons to generate executable assertions together with the computation graphs and tensor inputs they check.
Given a target compiler, the agent selects a compatible combination of an operator, a property, and a data generator from the available choices, and then supplies concrete tensor parameters, compiler API calls, and output comparisons.
For example, an associativity check for \codefont{relax.add} compares both grouping orders on the same tensor inputs under the skeleton's shape, value, and comparison rules.
Other properties follow the same steps with their own skeletons' rules, such as reduction decomposition for \codefont{relax.sum} and idempotence for \codefont{relax.nn.relu}.

\begin{algorithm}[t]
\caption{Workflow of \tool.}
\label{alg:workflow}
\SetKwInOut{Input}{Input}\SetKwInOut{Output}{Output}
\SetCommentSty{textit}
\Input{Target compiler $C$, property skeletons $P$, data generators $G$, prioritization rules $R$, time budget $B$}
\Output{Recorded failures $A$}
$O \gets \textsc{DiscoverOperators}(C)$\tcp*[r]{identified operators}\label{alg:discover}
$U \gets \emptyset$\tcp*[r]{usage records}
$F \gets [\,]$\tcp*[r]{recent feedback}
$A \gets \emptyset$\tcp*[r]{recorded failures}
\While{time budget $B$ remains}{
    $d, s \gets \textsc{Instantiate}(O, P, G, R, U, F)$\tcp*[r]{decision record $d$, test script $s$}\label{alg:instantiate}
    $v \gets \textsc{StaticChecks}(d, s, O, P, G)$\tcp*[r]{validation verdict $v$}\label{alg:static}
    \lIf{$v$ accepts}{$v \gets \textsc{AgentReview}(d, s, P, G)$}\label{alg:review}
    \If{$v$ rejects}{\label{alg:reject-if}
        append the rejection reason to $F$; \textbf{continue}\;\label{alg:reject}
    }
    $e \gets \textsc{Execute}(C, s)$\tcp*[r]{execution result $e$}\label{alg:execute}
    update $U$ with $d$; append a summary of $e$ to $F$\;\label{alg:update}
    \lIf{$e$ is unsuccessful}{add the failure $(d, s, e)$ to $A$}\label{alg:bundle}
}
\Return{$A$}\;
\end{algorithm}

Second, \tool validates each candidate before execution to reduce errors in oracle instantiation.
Validation reviews property applicability, the generated computation graph and its tensor inputs, and the oracle's comparison rules.
It also screens the test for unsafe operations.
For example, a candidate can build a valid graph yet incorrectly require \codefont{relax.subtract} to be commutative, so its assertion would reject correct outputs when the operands differ.
Validation checks this applicability condition and returns rejection reasons as feedback for subsequent generation.
Only accepted candidates proceed to execution.

Third, \tool executes the validated tests and uses their outcomes to guide later generation.
For each executed test, \tool records the outcome and updates execution counts for operators, properties, data generators, and their combinations.
These records guide the agent toward unused or under-used combinations, while rejection feedback helps it avoid repeating unsuitable choices.

\MyPara{Evaluation.}
We evaluate \tool on three DL compilers (\ie TVM~\cite{tvm}, ONNX-MLIR~\cite{onnx-mlir}, and GeneSys~\cite{genesys}) against \agentonly, which uses the same agent setup as \tool but without its property skeletons, test validation, or feedback loop.
We run each approach with six agents, three proprietary (GPT-5.5, GPT-6.1, and Opus-5.5) and three open-weight (Qwen-3.8, DeepSeek-4.1, and MiniMax-3.1), for six hours on each compiler.
The results show three main findings.
First, \tool detects more compiler faults, including silent wrong outputs.
Its tests expose 81 distinct failure symptoms, 6.23$\times$ those of \agentonly, which trace back to 96 distinct root causes, 7.38$\times$ those of \agentonly.
Of these root causes, 25 produce silent wrong outputs, but \agentonly detects none of them.
In issue replication, \tool detects 2.63$\times$ the faulty TVM versions that \agentonly detects, and still 1.89$\times$ after the outdated API calls in the tests of \agentonly are repaired.
Second, \tool generates valid tests with every agent.
Its validity reaches 94.91\% across all generated tests, 22.82 percentage points above \agentonly, because skeletons constrain each test and validation and feedback catch the remaining errors, which reduces the invalid tests from 989 to 145.
Third, \tool generates more diverse tests.
With the same agent, its valid tests exercise up to 9.50$\times$ the operator--property--generator combinations of \agentonly, and combining the six agents, they cover at least 2.73$\times$ the operator--property pairs of \agentonly on each compiler.

In summary, this paper makes three contributions.
\begin{itemize}[leftmargin=*]
    \item We identify oracle construction as an under-explored gap in DL compiler testing and characterize what turning a tensor-algebraic relation into an executable check requires: applicable operators, valid computation graphs, matching tensor inputs, and precision-aware comparisons.
    \item We introduce reusable property skeletons and an agentic workflow that combines controlled instantiation, candidate validation, and feedback to construct executable semantic oracles.
    \item We evaluate \tool on three DL compilers with six agents against \agentonly.
    Results show that \tool exposes more compiler faults, including silent wrong outputs that \agentonly misses, detects more faulty TVM versions, keeps test validity high across agents, and improves test diversity.
\end{itemize}

\section{Background}
\label{sec:background}
\MyPara{Oracles in DL Compiler Testing.}
A test oracle determines whether an observed execution satisfies the expected behavior.
In DL compiler testing, crash-based oracles flag unexpected failures during compilation or execution~\cite{tzer,hirgen}.
Interpreting a failure also requires knowing whether the computation graph is valid.
For example, HirGen~\cite{hirgen} expects graphs that violate type or shape constraints to raise the corresponding exceptions and treats segmentation faults as compiler bugs.
These checks detect compilation and error-handling failures, but do not establish whether the compiled computation produces correct results.
To check numerical correctness, differential oracles compare compiled outputs with those from a reference implementation or another backend~\cite{nnsmith,neuri,hirgen}.

\MyPara{Tensor-Algebraic Properties as Strong Semantic Oracles.}
Tensor algebra provides strong semantic oracles by expressing correctness as algebraic relations between execution results~\cite{tensor-decomposition-textbook,polyjuice}.
For example, commutativity requires \codefont{add(x, y)} to agree with \codefont{add(y, x)} for compatible tensor inputs.
Idempotence requires \codefont{relu(relu(x))} to agree with \codefont{relu(x)}.
A reduction such as \codefont{sum(x)} can be checked by splitting \codefont{x}, reducing each part, and recomposing the partial sums.
These oracles can expose semantic errors that crash-based checks miss, without requiring an expected output to be written for every concrete tensor input.

However, an algebraic relation alone does not define an executable oracle.
Constructing such an oracle requires selecting applicable operators, constructing well-formed computation graphs and matching tensor inputs, and implementing the output comparison.
For example, commutativity applies to \codefont{add} and \codefont{multiply}, but not to \codefont{subtract}, while reduction decomposition requires a valid split axis and a recomposition rule.
The comparison must also account for the tensor values and numerical precision.
Regrouping floating-point additions can change rounded results, so exact equality may reject a correct reduction even when the mathematical identity holds.
These requirements must remain consistent as the relation is instantiated for different operators and compilers.
In this work, we use property skeletons to record algebraic relations together with their applicability conditions, tensor input constraints, and comparison rules.
These specifications guide controlled instantiation of executable assertions together with the computation graphs and tensor inputs they check.

\section{\large \tool}
\label{sec:approach}
\begingroup
\definecolor{skeletonaccent}{HTML}{991F2B}
\definecolor{skeletonslotfill}{HTML}{F0A6AC}
\definecolor{skeletoncomment}{HTML}{1A4A8A}
\newcommand{\skeletonfont}{\fontfamily{lmtt}\fontsize{8}{10}\selectfont}
\newcommand{\skeletonslot}[1]{%
  \begingroup
  \setlength{\fboxsep}{0.7pt}%
  \colorbox{skeletonslotfill}{%
    \skeletonfont\mdseries\color{skeletonaccent}\vphantom{Ag}\textless#1\textgreater}%
  \endgroup}
\begin{lstlisting}[
language=Python,
float=t,
caption={An example skeleton for decomposition identity. The agent fills the highlighted slots.},
label={lst:decomposition-skeleton},
captionpos=b,
basicstyle=\skeletonfont\color{black},
lineskip=0.8pt,
columns=fullflexible,
keepspaces=true,
numbers=none,
frame=tb,
framerule=0.35pt,
rulecolor=\color{black!40},
xleftmargin=0pt,
framexleftmargin=0pt,
linewidth=\linewidth,
keywordstyle=\bfseries\color{black},
otherkeywords={},
morekeywords={data,apply},
identifierstyle=\color{black},
commentstyle=\color{skeletoncomment},
escapeinside={(*@}{@*)}
]
# decomposition_identity
# Intent: op(x) == rewrite(x), where rewrite composes simpler operators.
# Applicability conditions: Exact rewrite; available operators; same shape, dtype, and broadcasting.
# Rejection rule: No exact rewrite is known, or the identity needs unstated preconditions.
# Oracle notes: Exact equality for exact arithmetic; tolerances for floating-point rewrites.
@data(tensors = (*@\skeletonslot{generator}@*)((*@\skeletonslot{shapes}@*), (*@\skeletonslot{dtype}@*), (*@\skeletonslot{values}@*)))
@apply(lhs = lambda xs: (*@\skeletonslot{op}@*)(*xs), rhs = lambda xs: (*@\skeletonslot{rewrite}@*)(*xs))
@assert(relation = EquivalentOutputs(oracle=(*@\skeletonslot{compare}@*)))
\end{lstlisting}
\endgroup

\tool constructs semantic oracles through controlled instantiation of property skeletons.
Its design addresses the gap between a mathematical relation and an executable check.
An executable check also needs to know which operators a relation applies to, which tensors it holds on, and how to compare floating-point outputs.
\tool provides this knowledge through two insights.
First, these conditions can be reused across operators and compilers, so \tool records them once and applies them to every generated test.
Second, given these recorded conditions, an LLM agent can write the compiler-specific parts of an oracle, such as API calls and concrete tensors, and a separate review can check the result.

Algorithm~\ref{alg:workflow} outlines the workflow of \tool.
After \tool identifies the operators of the target compiler (Line~\ref{alg:discover}), each iteration constructs one oracle through three stages, as shown in Figure~\ref{fig:workflow}.
The agent first instantiates a property skeleton into executable assertions together with the computation graphs and tensor inputs they check (Line~\ref{alg:instantiate}).
Static checks and an agent review then decide whether the candidate proceeds (Lines~\ref{alg:static}--\ref{alg:reject}).
Finally, \tool executes the accepted candidate and records its outcome to guide the next iteration (Lines~\ref{alg:execute}--\ref{alg:bundle}).
All three stages build on the skeletons and data generators, so we introduce them first in Section~\ref{sec:property-skeletons}.
We then describe the stages in Sections~\ref{sec:controlled-generation}, \ref{sec:validation}, and \ref{sec:execution-feedback}.

\subsection{Reusable Property Skeletons}
\label{sec:property-skeletons}

An algebraic relation provides an oracle only when its assumptions hold for the tested computation.
For example, commutativity states that swapping two operands preserves the result~\cite{tensor-decomposition-textbook}, but this relation applies to addition and not subtraction.
Tensor shapes, values, and numerical precision can impose further conditions, since regrouped floating-point additions can round differently.
\tool records these conditions together with the relation in a reusable document called a property skeleton, so the agent can consult them for every test.

\MyPara{Skeleton Structure.}
Listing~\ref{lst:decomposition-skeleton} shows an example skeleton for decomposition identity, which compares an operator with an equivalent composition of simpler operators.
The equation \codefont{op(x) == rewrite(x)} looks simple, yet writing a valid oracle from it requires knowledge that the equation does not state.
The rewrite must use operators that the target compiler provides and must preserve shapes, data types, and broadcasting behavior.
It must also hold without hidden preconditions, so rewriting \codefont{sqrt(x * x)} as \codefont{x} is invalid because the two sides differ for negative \codefont{x}.
Even a correct rewrite needs a suitable comparison, since exact arithmetic permits exact equality whereas floating-point rewrites need tolerances.
The skeleton states this knowledge in natural language through its intent, applicability conditions, rejection rule, and oracle notes, which both the generation agent and the validation agent consult.

\begin{table*}[t]
\centering
\caption{The 16 property skeletons in three categories: 11 algebraic properties, two tensor properties, and three fallback properties for operators that no algebraic or tensor relation covers.
    Each row gives an example relation or check and the applicability conditions under which it is a valid oracle.}
\label{tab:property-taxonomy}
\footnotesize
\setlength{\tabcolsep}{3pt}
\renewcommand{\arraystretch}{1.02}
\begin{tabular}{@{}>{\raggedright\arraybackslash}p{0.095\textwidth}c>{\raggedright\arraybackslash\hyphenpenalty=10000}p{0.18\textwidth}>{\raggedright\arraybackslash}p{0.27\textwidth}>{\raggedright\arraybackslash}p{\dimexpr0.455\textwidth-8\tabcolsep-24pt\relax}@{}}
\toprule
\textbf{Category} & \textbf{Count} & \textbf{Property} & \textbf{Example Relation or Check} & \textbf{Applicability Conditions} \\
\midrule
\multirow[t]{11}{=}{Algebraic\newline Properties}
& \multirow[t]{11}{*}{11}
& Commutativity
& $\operatorname{max}(x,y)=\operatorname{max}(y,x)$
& Compatible shapes; finite values \\
& & Associativity
& $x+(y+z)=(x+y)+z$
& Matching shapes; appropriate tolerance \\
& & Identity
& $x\times 1=x$
& Representable identity; valid broadcasting \\
& & Idempotence
& $\operatorname{relu}(\operatorname{relu}(x))=\operatorname{relu}(x)$
& Valid tensors for repeated application \\
& & Cancellation
& $(x+y)-y=x$
& Supported inverse pair; stable value domain \\
& & Distributivity
& $x\times(y+z)=x\times y+x\times z$
& Compatible shapes; exact arithmetic \\
& & Decomposition Identity
& $\operatorname{ewise\_fma}(a,b,c)=a\times b+c$
& Available component operators;\newline matching shapes; exact arithmetic \\
& & Monotonicity
& $x\leq y \Rightarrow \operatorname{relu}(x)\leq\operatorname{relu}(y)$
& Ordered tensors in a monotone domain \\
& & Positive Homogeneity
& $\operatorname{relu}(\alpha x)=\alpha\operatorname{relu}(x)$
& $\alpha\geq 0$; representable scaled values \\
& & Shift Invariance
& $\operatorname{softmax}(x+c)=\operatorname{softmax}(x)$
& Uniform offset $c$ along normalization axis;\newline appropriate tolerance \\
& & Mathematical Semantics Consistency
& $x\equiv y \Rightarrow C(x)\equiv C(y)$
& Equality model permits substitution\newline in observation context $C$ \\
\midrule
\multirow[t]{2}{=}{Tensor\newline Properties}
& \multirow[t]{2}{*}{2}
& Permutation Invariance
& $\operatorname{sum}(x)=\operatorname{sum}(\operatorname{permute}(x))$
& Permutation of reduced elements;\newline appropriate tolerance \\
& & Geometric Decomposition
& $x+y=\operatorname{concat}(x_1+y_1,x_2+y_2)$
& Matching splits and concatenation axis \\
\midrule
\multirow[t]{3}{=}{Fallback\newline Properties}
& \multirow[t]{3}{*}{3}
& Reference Consistency
& Compiled vs. reference outputs
& Matching semantics; appropriate tolerance \\
& & Concrete Input Replay
& Valid example across compilation settings
& Same graph and tensors; supported settings \\
& & Shape and Dtype Contract
& Output shape and dtype vs. specification
& Known output shape and dtype \\
\bottomrule
\end{tabular}

\end{table*}

A skeleton also ties the generated code to the intended relation, so that a test keeps comparing the rewrite under the specified comparison.
\tool controls instantiation through the last three lines of Listing~\ref{lst:decomposition-skeleton}, which form the skeleton's oracle template.
The template fixes how a test draws tensor inputs (\codefont{@data}), computes both sides of the relation (\codefont{@apply}), and compares their outputs (\codefont{@assert}), and it leaves the highlighted slots for the agent to fill.
Some slots take functions and others take plain values: \codefont{<generator>} takes a data generator, \codefont{<op>} and \codefont{<rewrite>} take the two computations, and \codefont{<compare>} takes a comparison, while \codefont{<shapes>}, \codefont{<dtype>}, and \codefont{<values>} take plain values.
The template thus works like a higher-order function, and instantiating a skeleton amounts to applying it to arguments written for the target compiler.
As a result, the agent chooses the arguments, while the skeleton preserves the relation between them.

The same template format also serves the other skeletons, because the two sides in \codefont{@apply} can be any computations on the tensor inputs.
In Listing~\ref{lst:decomposition-skeleton}, the two sides apply the operator and its rewrite to the same tensor inputs, and other skeletons build the two sides in other ways.
Associativity nests calls, as in \codefont{op(op(x,y),z)} versus \codefont{op(x,op(y,z))}.
Permutation invariance applies the operator to a reordered input tensor, as in \codefont{sum(x)} versus \codefont{sum(permute(x))}.
Geometric decomposition applies the operator to parts of the input tensor and recombines the partial results, as in \codefont{sum(x)} versus \codefont{sum(x1) + sum(x2)}.
Cancellation applies an operation and its inverse with a helper operand, as in \codefont{(x+w)-w} versus \codefont{x}.
Likewise, the comparison in \codefont{@assert} can be exact equality, equality within a tolerance, or an order relation, such as \codefont{relu(x) <= relu(y)} for monotonicity.

\MyPara{Skeleton Inventory.}
DL compiler operators satisfy different relations, so \tool provides 16 property skeletons in three categories, listed in Table~\ref{tab:property-taxonomy}.
Each row pairs a representative oracle in the Example Relation or Check column with the assumptions it requires in the Applicability Conditions column.

The 11 algebraic properties state relations about operators and their compositions, and what each relation changes determines the conditions it imposes.
Commutativity, associativity, and distributivity rearrange operands, as in $x+y$ and $y+x$, so they require compatible shapes across the rearranged computations.
Identity, cancellation, and idempotence instead relate an operator to an identity element, an inverse, or its own repeated application, so their conditions restrict tensor values.
For example, the identity element must be representable in the selected data type, and multiplicative cancellation must avoid divisors near zero.
Decomposition identity relates a compound operator to its components, such as \codefont{ewise\_fma(a,b,c)} and $a\times b+c$.
Monotonicity, positive homogeneity, and shift invariance describe how an output changes when the input tensor is increased, scaled, or shifted, so their conditions restrict that change.
For example, \codefont{softmax} ignores a uniform offset only along its normalization axis, while \codefont{relu} scales with its input tensor only when $\alpha\geq 0$.
Finally, mathematical semantics consistency asks whether values that compare as equal remain interchangeable.
For instance, \codefont{+0.0} and \codefont{-0.0} compare as equal, yet their reciprocals have opposite signs, so this skeleton requires each test to state which notion of equality it checks.

Some operators also satisfy relations about how tensor elements are arranged, and the two tensor properties capture these relations by reordering or partitioning the elements.
Permutation invariance requires an operator that ignores element order, such as a reduction over all elements.
Geometric decomposition splits a tensor along an axis, applies the operator to the whole tensor and to each part, and recomposes the partial results.
The correct recomposition depends on the axes, so its skeleton states a rule for each case.
When the split and reduction axes differ, the partial results are concatenated.
When they match, the combiner depends on the operator: \codefont{sum} adds the partial sums, while \codefont{mean} takes their average weighted by the split sizes.
Its splits also combine with other properties, so one test can check two properties at once.
For example, \codefont{op(x, y) == concat(op(y1, x1), op(y2, x2))} checks the decomposition and commutativity together.

The three fallback properties extend testing to operators without an algebraic or tensor relation.
Reference consistency compares compiled outputs with a reference implementation such as NumPy.
When no reference is available, shape and dtype contract compares the output shape and data type with their expected values, which suits operators such as \codefont{reshape} and \codefont{concat}.
Finally, concrete input replay executes a valid example across compilation settings, which detects failures such as compiler crashes and exceptions.
Because each skeleton is a self-contained document, this inventory can be readily extended, and supporting a new property only requires writing its document.

\begin{figure}[t]
\centering
\includegraphics[width=\linewidth]{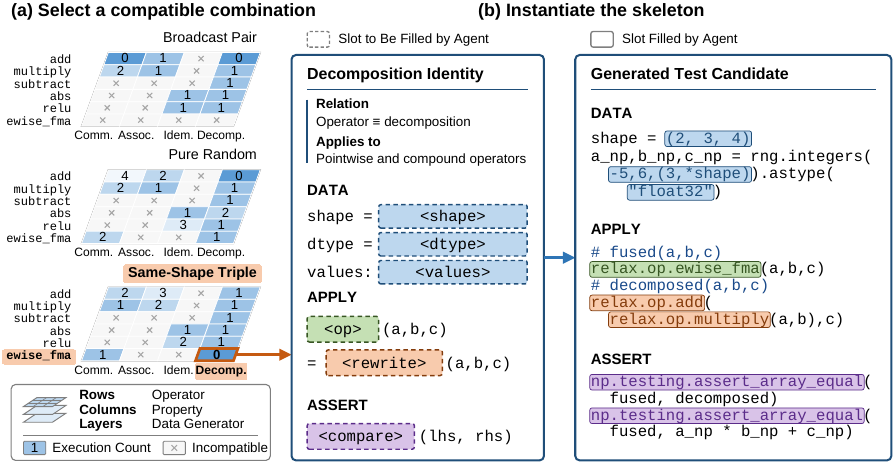}
\caption{Controlled oracle instantiation for TVM's fused multiply-add operator \codefont{ewise\_fma}.
(a) Oracle selection. The stacked matrices index operators by row, properties by column, and data generators by layer.
Each cell shows the execution count of its combination, a cross marks an incompatible combination, and darker cells have fewer executions and receive higher priority.
\tool selects an unexecuted combination of \codefont{ewise\_fma}, decomposition identity, and Same-Shape Triple.
(b) Oracle instantiation. The agent fills the skeleton's slots with input tensors, the fused and decomposed computations, and output comparisons, and matching colors connect each slot to its generated code.}
\label{fig:approach-instantiation}
\end{figure}

\MyPara{Data Generators.}
An oracle also needs tensor inputs that make its relation meaningful.
For example, monotonicity compares outputs for tensors ordered by $x\leq y$.
\tool therefore separates the choice of tensors from the choice of relation: a property defines the relation to check, while a data generator determines the tensor inputs used to exercise it.
These tensors are the values that flow through a computation graph, and the agent builds the graph itself from the selected operator and property.
This separation lets one generator serve different relations, as Same-Shape Triple serves both associativity and decomposition identity.
Like a skeleton, each generator is a document that states its intent, suitable uses, applicability conditions, and rejection rule, and it leaves parameters open for the agent.
For example, the Same-Shape Triple document leaves the shape, data type, and value range for the agent to fill.
The 20 data generators of \tool form five families.
\begin{itemize}[leftmargin=*]
    \item {\bf Random Sampling.}
    Some relations hold on arbitrary values, so their tensors can be drawn at random.
    \emph{Pure Random} draws tensors with a fixed seed and no further constraint, such as two random tensors for checking \codefont{add(x, y) == add(y, x)}.
    \emph{Domain-Aware Random} also draws random tensors but keeps them in an operator's value domain, such as positive tensors for \codefont{relax.log}.
    \item {\bf Value Regimes.}
    A compiler can mishandle specific values that random sampling rarely draws, so these generators place such values in the tensors.
    \emph{Special Values} uses small constants where an operator's behavior changes, such as $0$ and $\pm10^{-6}$ for \codefont{relu}.
    \emph{Data Type Boundaries} uses the extreme values of a data type, such as $-128$ and $127$ for \codefont{int8}, to exercise overflow.
    \emph{Reference Edge Cases} uses values on which implementations can disagree, such as $2.5$ for \codefont{round}, which NumPy rounds to $2$.
    \emph{Semantic Gap Witnesses} uses values that compare as equal yet produce different results, such as \codefont{+0.0} and \codefont{-0.0}, whose reciprocals have opposite signs.
    \emph{Reduction Stress} mixes large and small magnitudes, such as $[10^{8},10^{-8},-10^{8}]$, whose floating-point sum depends on the order of additions.
    \item {\bf Tensor Relations.}
    Some properties hold only for related tensors, and independent sampling rarely produces them, so these generators construct the tensors together.
    \emph{Same-Shape Triple} provides three tensors with the same shape and data type, as \codefont{ewise\_fma(a, b, c)} requires.
    \emph{Duplicated Tensor} uses one tensor twice, as in \codefont{maximum(x, x) == x}.
    \emph{Ordered Pair} provides two tensors with $x\leq y$ at every element, such as $[1,2]$ and $[1,5]$.
    \emph{Identity Injection} pairs a tensor with an operator's identity element, such as a zero tensor in \codefont{add(x, 0) == x}.
    \emph{Cancellation Witness} pairs a tensor with an operand that can be undone, such as a nonzero $w$ in \codefont{(x * w) / w == x}.
    \item {\bf Tensor Structure.}
    Some checks depend on how tensors are shaped or partitioned, so these generators fix this structure and record it for the assertions.
    \emph{Broadcast Pair} provides two tensors with broadcast-compatible shapes, such as $(2,3)$ and $(1,3)$, and records the expected result shape $(2,3)$.
    \emph{Permuted Tensor} provides a tensor and a reordered copy, such as $[1,2,3]$ and $[3,1,2]$, and records the permutation.
    \emph{Axis Split} divides a tensor along an axis, such as a $(4,3)$ tensor into $(1,3)$ and $(3,3)$ parts, and records how to recombine them.
    \emph{Shape Edge Cases} uses shapes that can expose shape-handling bugs, such as a rank-0 scalar or a $(1,3,1)$ tensor.
    \emph{Axis and Index Attributes} provides valid arguments for axis-dependent operators, such as \codefont{axis=-1} and \codefont{keepdims=True} for \codefont{sum}.
    \emph{Reshape-Compatible Shapes} pairs a tensor with a target shape of the same element count, such as $(3,4)$ and $(2,6)$.
    \item {\bf Signature-Driven Tensors.}
    The fallback properties need operator calls that run, so these generators read an operator's signature and supply a valid tensor or attribute for each required argument.
    \emph{Contract Tensors} uses arguments whose output shape and data type can be computed in advance, such as $(2,3)$ and $(3,4)$ matrices for \codefont{matmul}, which must return a $(2,4)$ matrix.
    \emph{Smoke Tensors} uses small conservative arguments when the expected output is unknown, such as a $(2,2)$ tensor of ones, so the test checks only that the operator compiles and runs.
\end{itemize}
New data generators can be added in the same way, since each new generator document becomes available to existing skeletons.

\begin{figure}[t]
\centering
\includegraphics[width=\linewidth]{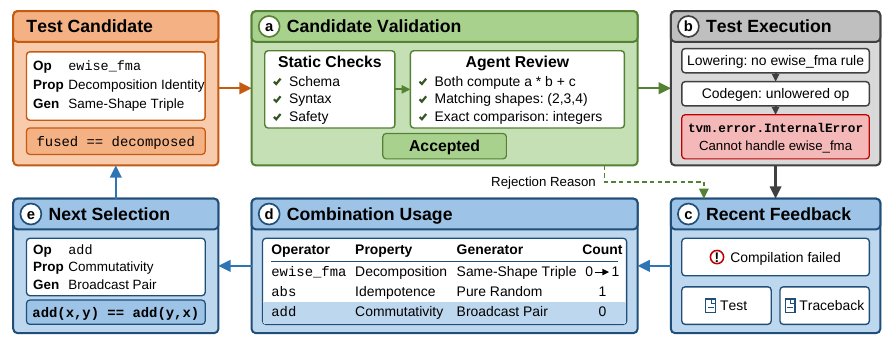}
\caption{Validation, execution, and feedback for the \codefont{ewise\_fma} test of Figure~\ref{fig:approach-instantiation}.
(a) The candidate passes static checks on its schema, syntax, and safety, and an agent review confirms that both computations evaluate $a \times b + c$ on matching shapes with integer-valued inputs that permit exact comparison.
(b) Execution exposes a TVM compiler bug: TVM provides \codefont{ewise\_fma} but cannot lower it, so code generation fails with an internal error.
(c) \tool records the failure as recent feedback, and (d) it increments the execution count of the selected combination.
(e) Usage balancing then directs the next selection to the unexecuted combination of \codefont{add}, commutativity, and Broadcast Pair.
A rejected candidate would instead skip execution, and the dashed path would carry its rejection reason into the feedback.}
\label{fig:approach-feedback}
\end{figure}

\subsection{Controlled Oracle Instantiation}
\label{sec:controlled-generation}

Skeletons and data generators specify what a valid oracle requires, and the agent writes this oracle for a concrete compiler, including its executable assertions, computation graphs, and tensor inputs.
\tool controls how the agent constructs each oracle in two steps, in which the agent first selects the oracle to instantiate and then fills the slots of its skeleton.

\MyPara{Step 1. Oracle Selection.}
Each oracle starts from an operator--property--generator combination, because these three choices together determine what the oracle checks.
The property supplies the relation, the operator determines where the relation is checked, and the data generator determines the tensor inputs that exercise it.
To form these combinations, \tool identifies the operators to test before the first iteration, using the target compiler's registered operators or source code.
In our experiments, we observe that \tool built on top of GPT-6.1 identifies 175 operators for TVM.
With 16 skeletons and 20 data generators, these 175 TVM operators yield 56,000 combinations before applicability conditions exclude the incompatible ones.

To choose within a space of this size, the agent needs the conditions of each property and data generator, as well as a limit on what it may select.
The generation prompt therefore lists the identified operators, includes the complete skeleton and data generator documents, and asks the agent to follow their constraints.
The agent must select its operator, property, and data generator from these lists, so every oracle can be traced to a documented relation.
Within that selection, it can freely choose shapes, data types, axes, tolerances, seeds, and value ranges, because these choices depend on the operator and the compiler.
The prompt also supplies recent feedback, usage records, and prioritization rules to guide the selection, which we will discuss in Section~\ref{sec:execution-feedback}.

Figure~\ref{fig:approach-instantiation}(a) shows this selection for six operators, four properties, and three data generators.
Its stacked matrices organize operators by row, properties by column, and data generators by layer, and a cross marks an incompatible combination.
For example, \codefont{subtract} has a cross under commutativity, and \codefont{ewise\_fma} has crosses throughout the Broadcast Pair layer because TVM requires its three operands to have equal shapes.
Each compatible cell shows an execution count, and darker cells have fewer prior executions and receive higher priority.
In this example, \tool selects TVM's fused multiply-add operator \codefont{ewise\_fma}, decomposition identity, and Same-Shape Triple, a combination that has not yet been executed.

\MyPara{Step 2. Oracle Instantiation.}
Once the combination is selected, the agent supplies the skeleton's arguments and thereby turns its relation into executable assertions on the target compiler.
These assertions are meaningful only on a computation graph and tensor inputs that satisfy the relation's assumptions, so the agent constructs the three jointly.
The tensor inputs must be valid for both computations, so an oracle for \codefont{relax.divide} needs a nonzero divisor.
The computations in turn determine the comparison, so a floating-point regrouping requires a tolerance.
The comparison also constrains the tensor inputs, so an exact-equality assertion requires tensor values for which the compared computations produce the same representable result, such as small integers.

To make these choices reviewable and reproducible, the agent returns two artifacts for each oracle.
The first is a decision record that lists the selected combination, its parameters such as shapes and data types, its expected oracle (\ie\ the outcome that its assertions should observe), and a test scenario whose \codefont{data}, \codefont{apply}, and \codefont{assert} fields mirror the skeleton's annotations.
The second is a self-contained Python script that implements the scenario with a deterministic seed.
This script also adapts the oracle to the compiler's interface.
A TVM script builds Relax functions and invokes the compiler directly, whereas an ONNX-MLIR or GeneSys script creates a small static-shape ONNX model~\cite{onnx} that \tool then passes to the compiler.
As a result, a single commutativity skeleton can yield oracles for \codefont{relax.add} on TVM and for \codefont{Add} nodes on ONNX-MLIR and GeneSys.

\MyPara{Example.}
Figure~\ref{fig:approach-instantiation}(b) shows how \tool instantiates the selected skeleton, where matching colors connect each dashed slot to its generated code.
The blue slots receive the tensor inputs from Same-Shape Triple, which are three \codefont{float32} tensors of shape $(2,3,4)$ with integer values between $-5$ and $5$.
The green and orange slots receive the two computations, so \tool builds two Relax functions, \codefont{fused} and \codefont{decomposed}, that compute \codefont{ewise\_fma(a, b, c)} and \codefont{add(multiply(a, b), c)} on the same tensor inputs.
Finally, the purple slot receives the comparison.
These small integer products and sums are exactly representable, so \tool uses \codefont{np.testing.assert\_array\_equal} to require the compiled outputs to match exactly.
It also compares the fused output with NumPy's evaluation of \codefont{a * b + c} and checks the output shapes and data types.

\subsection{Candidate Validation}
\label{sec:validation}

A skeleton bounds the agent's choices, and validation confirms that a test's assertions check the selected oracle.
For example, a candidate may select the shape and dtype contract for ONNX's \codefont{Conv} operator on ONNX-MLIR, where \tool compiles the model, so its assertions must observe the output shape and data type.
\tool validates each candidate before execution in two stages, as shown in Figure~\ref{fig:approach-feedback}(a).
Static checks first filter out malformed candidates, and an agent then reviews the semantics of the remaining ones.

\MyPara{Stage 1. Static Checks.}
Some errors require no semantic judgment, so the first stage detects them through deterministic checks that need no agent call.
\begin{itemize}[leftmargin=*]
    \item {\bf Schema Checks.}
    Later stages rely on the decision record, so \tool checks that the scenario contains nonempty \codefont{data}, \codefont{apply}, and \codefont{assert} fields.
    It also verifies that the operator, property, and data generator belong to the provided lists and that the decision records a seed and expected oracle.
    \item {\bf Syntax and Target Checks.}
    A test must be able to run and reach the compiler, so \tool compiles the Python script to detect syntax errors.
    It also checks that a TVM script imports TVM and that an ONNX-MLIR or GeneSys script imports ONNX and saves \codefont{model.onnx}.
    \item {\bf Safety Checks.}
    \tool executes generated code, so it parses each script and screens for forbidden calls and imports, including dynamic code execution, subprocess creation, and network access.
    It also screens for filesystem writes outside the permitted output files.
\end{itemize}

\MyPara{Stage 2. Agent Review.}
Judging whether a test is semantically appropriate requires reasoning about its intent, so the second stage relies on a validation agent, which judges the decision and script against the documents of the selected property and data generator.
This agent is separate from the generation agent, so it reviews the candidate without the generation conversation.
Its report covers four checks and records evidence for each one.
\begin{itemize}[leftmargin=*]
    \item {\bf Operator Plausibility.}
    An oracle is useful only if its test exercises the selected operator, so the script must plausibly reach that operator on the target compiler.
    For example, the review rejects a call to \codefont{relax.op.nn.pixel\_shuffle} with a \codefont{layout} keyword, which this TVM API does not accept.
    \item {\bf Property and Generator Applicability.}
    A relation is a valid oracle only under its assumptions, so it must apply to the operator and tensor inputs, and the data generator must suit the property.
    For example, the review rejects shift invariance for \codefont{Cos} because cosine repeats only at multiples of $2\pi$.
    It also rejects Special Values for \codefont{IsNaN} because every value in the finite pool yields false, so the value choice has no effect.
    \item {\bf Scenario Consistency.}
    The decision record states the intended oracle, so the script must implement this scenario.
    For example, the review rejects a candidate for \codefont{relax.einsum}, an operator that takes a subscript string such as \codefont{ij,jk->ik} to describe a tensor contraction, because its decision record lists seven such strings while its script tests eight, so the record no longer describes what the script checks.
    \item {\bf Script Safety.}
    Static screening matches fixed patterns, so the agent also reviews whether the script appears to write outside its permitted files or reach external services.
\end{itemize}

A rejected candidate skips execution, and the review returns a concise reason that guides the next generation step.
This reason can name a concrete repair, so the agent can fix the mistake while keeping the parts of the test that were correct.
For example, the review of the \codefont{pixel\_shuffle} candidate asks the agent to drop the unsupported \codefont{layout} keyword, and the next candidate makes only this change and passes the review.

\MyPara{Example.}
The fused multiply-add test passes both stages, as shown in Figure~\ref{fig:approach-feedback}(a).
Its decision record is complete, and its script is well-formed and free of forbidden operations.
The reviewer then checks that the rewrite computes the same expression, the three input tensors have matching shapes, and the small integer values make exact comparison appropriate.

\subsection{Test Execution and Feedback}
\label{sec:execution-feedback}

\tool executes each accepted candidate on the target compiler to expose compiler faults and feeds the outcome back into subsequent instantiation.

\MyPara{Test Execution.}
Generated tests can crash or hang, so \tool runs each test in an isolated process with a time limit.
How a test reaches the compiler then depends on the compiler's interface.
A TVM test invokes the compiler within its Python script, so \tool classifies the script's result as a satisfied property, a counterexample from a failed assertion, a native crash, or another exception.
A GeneSys or ONNX-MLIR test instead generates an ONNX model.
\tool first validates this model with the ONNX checker and shape inference, so it can distinguish a malformed model from a compiler failure.
\tool then compiles the model and reports a build error, an unsupported operator or shape, or a timeout.

\MyPara{Failure Artifacts.}
A reported failure can come from the compiler or from the test itself, so each failure must be reproducible and open to review.
For each unsuccessful execution, \tool therefore saves the selected combination, the traceback, a reproduction command, the seed, and the recipe for the tensor inputs.
A separate agent then assigns a short failure label, such as crash or numerical instability, which supports later triage.

\MyPara{Feedback Signals.}
Feedback tells each generation step which choices to avoid and which oracles already exist, through three signals in the next generation prompt.
Rejection feedback carries the reason that the review gives for each rejected candidate (Section~\ref{sec:validation}), so the agent can avoid repeating the mistake.
An execution summary reports the outcome, the selected combination, the compilation phase, and the top exception, which helps the agent repair a faulty test or explore around a failure.
Usage records count how often executed tests use each operator, property, data generator, and their combinations, so the agent can see which of them remain untested.
To keep the prompt bounded as testing proceeds, \tool keeps each summary concise and includes only the five most recent feedback entries.

\MyPara{Prioritization Rules.}
Prioritization rules then use the rejection reasons, execution summaries, and usage records to tell the agent which combination to select next.
For example, after addition's commutativity has been checked with one data generator, the rules direct the agent toward other data generators.
These rules are inspired by coverage-guided testing~\cite{afl,libfuzzer}.
Coverage-guided fuzzers favor inputs that reach new code, whereas \tool favors combinations that no executed test has used.
\tool reads the combinations of executed tests from their decision records, so it needs no compiler instrumentation.
The agent applies two rules in order.
\begin{itemize}[leftmargin=*]
    \item {\bf Usage Balancing.}
    This rule asks the agent to first select operators, properties, and data generators that no executed test has used, so that testing spreads across the compiler.
    Once every operator has been used, the agent selects the operator with the fewest executions and pairs it with its least-used properties and data generators.
    \item {\bf Failure-Adjacent Exploration.}
    A prior failure suggests that related tests may expose further faults, so this rule then directs the agent toward prior counterexamples, crashes, timeouts, and repeated rejections.
    For example, if \codefont{relax.sum} with geometric decomposition and Reduction Stress reports a counterexample, the agent can try another reduction axis, split point, or related reduction operator.
\end{itemize}
These rules guide which oracle to instantiate next, while the selected skeleton constrains how to construct it.
Like a skeleton or a data generator, each rule is a document that \tool includes in the generation prompt, so adding a rule only requires writing its document.

\MyPara{Example.}
Figure~\ref{fig:approach-feedback} traces the fused multiply-add test after validation.
When \tool executes the script, TVM fails at \codefont{relax.build}, before either numerical comparison runs, as shown in Figure~\ref{fig:approach-feedback}(b).
TVM provides \codefont{ewise\_fma} as a Relax operator but has no rule that lowers it into lower-level code.
The operator therefore keeps its original form until code generation, where \codefont{CodeGenVM} fails on it with a \codefont{tvm.error.InternalError} instead of reporting it as unsupported.
As shown in Figure~\ref{fig:approach-feedback}(c), \tool saves the failing test and traceback and reports the failure and its top exception in the next generation prompt.
The execution also increments the selected combination's count from zero to one, as shown in Figure~\ref{fig:approach-feedback}(d).
In this table, \codefont{add} with commutativity and Broadcast Pair has zero executions, while the other two combinations have already been exercised.
Usage balancing therefore prioritizes this unused combination in Figure~\ref{fig:approach-feedback}(e), which checks that swapping two broadcast-compatible input tensors preserves the addition result.
The cycle then repeats with a new skeleton instantiation.

\section{Evaluation Results}
\label{sec:evaluation}
We seek to answer the following research questions.
\begin{description}
    \item[RQ1] How reliably does \tool generate valid tests, compared with \agentonly with the same agent?
    \item[RQ2] How diverse are the operators, properties, data generators, and their combinations used by valid tests generated by \tool, compared with \agentonly with the same agent?
    \item[RQ3] What compiler faults do tests generated by \tool detect, which of them does \agentonly miss with the same agent, and which checks detect them?
    \item[RQ4] How do property skeletons, validation, and feedback contribute to these results?
\end{description}

\begin{figure*}[t]
\centering
\input{figures/eval-style}
\pgfplotsset{table/col sep=comma}
\begin{tikzpicture}
\begin{groupplot}[group style={group name=val, group size=3 by 1, horizontal sep=9pt}, eval validity bars, height=120pt, ymax=6.55, ytick={1,...,6}, /pgf/bar width=6.4pt]
\nextgroupplot[title={TVM}, yticklabels={MiniMax-3.1,DeepSeek-4.1,Qwen-3.8,Opus-5.5,GPT-6.1,GPT-5.5}, xlabel={Test Validity (\%)}]
\draw[black!30, line width=0.3pt] (axis cs:0,3.5) -- (axis cs:100,3.5);
\addplot[eval bar=evalPropilot, bar shift=3.7pt, keep compiler=1] table[y=position, x=propilot_validity] {figures/data/rq1-validity.csv};
\addplot[eval bar=evalAgent, eval texture=evalNE, bar shift=-3.7pt, keep compiler=1] table[y=position, x=agent_only_validity] {figures/data/rq1-validity.csv};
\node[anchor=west, inner sep=1.2pt, font=\sffamily\fontsize{6}{7}\selectfont] at ([yshift=-3.7pt]axis cs:0,1) {No Test};
\nextgroupplot[title={ONNX-MLIR}, yticklabels={}, xlabel={Test Validity (\%)}]
\draw[black!30, line width=0.3pt] (axis cs:0,3.5) -- (axis cs:100,3.5);
\addplot[eval bar=evalPropilot, bar shift=3.7pt, keep compiler=2] table[y=position, x=propilot_validity] {figures/data/rq1-validity.csv};
\addplot[eval bar=evalAgent, eval texture=evalNE, bar shift=-3.7pt, keep compiler=2] table[y=position, x=agent_only_validity] {figures/data/rq1-validity.csv};
\nextgroupplot[title={GeneSys}, yticklabels={}, xlabel={Test Validity (\%)}]
\draw[black!30, line width=0.3pt] (axis cs:0,3.5) -- (axis cs:100,3.5);
\addplot[eval bar=evalPropilot, bar shift=3.7pt, keep compiler=3] table[y=position, x=propilot_validity] {figures/data/rq1-validity.csv};
\addplot[eval bar=evalAgent, eval texture=evalNE, bar shift=-3.7pt, keep compiler=3] table[y=position, x=agent_only_validity] {figures/data/rq1-validity.csv};
\end{groupplot}
\begin{scope}[shift={(val c1r1.north west)}, yshift=20pt]
\fill[evalPropilot] (0pt,-3pt) rectangle (12pt,3pt);\node[anchor=west, inner sep=0pt, font=\sffamily\scriptsize] at (14pt,0pt) {\toolname{}};
\path[fill=evalAgent, postaction={pattern=evalNE, pattern color=white}] (58pt,-3pt) rectangle (70pt,3pt);\node[anchor=west, inner sep=0pt, font=\sffamily\scriptsize] at (72pt,0pt) {\agentname{}};
\end{scope}
\end{tikzpicture}
\caption{Test validity of \tool and \agentonly.
\tool improves test validity by up to 90.07 percentage points.}
\label{fig:rq1-validity}
\end{figure*}

\MyPara{Compilers under Test.}
We evaluate \tool on the following three commonly used DL compilers.
\begin{itemize}[leftmargin=*]
    \item TVM~\cite{tvm}.
    This compiler deploys DL models onto platforms such as CPUs, GPUs, and specialized accelerators.
    We use TVM v0.23.0 at commit \codefont{75ce1b4}.
    \item ONNX-MLIR~\cite{onnx-mlir}.
    This compiler transforms a model into MLIR-based representations for further optimization on hardware backends.
    We use ONNX-MLIR v0.4.2 at commit \codefont{4a13c34}.
    \item GeneSys~\cite{genesys}.
    This compiler is one of the most recent research efforts in deploying DL models onto specialized hardware.
    We use GeneSys commit \codefont{8c5e488}.
\end{itemize}

\MyPara{Configurations.}
\tool is agent-agnostic.
We therefore build \tool on six recent agents to evaluate its efficacy across agents and run each configuration for six hours on each compiler.
\begin{itemize}[leftmargin=*]
    \item Proprietary agents.
    We build \tool on Codex~\cite{codex} with GPT-5.5 and GPT-6.1 and on Claude Code~\cite{claudecode} with Opus-5.5, which gives \propgptfive, \propgptsix, and \propopus.
    These LLMs are accessed through provider-managed CLI surfaces.
    \item Open-weight agents.
    We build \tool on Qwen Code~\cite{qwencode} with Qwen3.8-Flash-Next (Qwen-3.8), on DeepSeek Harness~\cite{deepseekharness} with DeepSeek-V4.1-Flash (DeepSeek-4.1), and on Codex~\cite{codex} with MiniMax-M3.1-Flash (MiniMax-3.1), which gives \propqwen, \propdeepseek, and \propminimax.
    These open-weight LLMs are served by the National Research Platform~\cite{nrp}.
\end{itemize}

\MyPara{Baselines.}
We compare \tool with eight baselines, which can be categorized into three groups.
\begin{itemize}[leftmargin=*]
    \item \agentonly.
    This approach uses the same agent, \ie the same harness and LLM, as the corresponding \tool configuration to generate PBTs, without its controlled tensor-knowledge-guided test generation, validation, or feedback loop, which gives \agentgptfive, \agentgptsix, \agentopus, \agentqwen, \agentdeepseek, and \agentminimax.
    \item \plusskeleton.
    This variant of \propminimax adds property skeletons to \agentminimax, without validation or feedback.
    \item \plusvalidation.
    This variant of \propminimax adds property skeletons and validation to \agentminimax, without feedback.
\end{itemize}

\subsection{RQ1: Test Validity}
\label{sec:rq1}
We first compare the validity of the tests generated by \tool and \agentonly with each agent.
A valid test must construct a well-formed compiler input and provide a correct oracle for an applicable semantic property.
Tests are invalid if they use an inapplicable property, misuse APIs, construct malformed computation graphs, contain syntax errors or undefined variables, require missing test fixtures, or have incorrect or unverifiable oracles.
We first check that a test executes, and we then verify its oracle in the declared precision on an independent reference, \ie NumPy for TVM and ONNX's \codefont{ReferenceEvaluator} for ONNX-MLIR and GeneSys.
When no suitable reference can verify an oracle, we conservatively classify the test as invalid.
A model that a compiler explicitly declares unsupported is also invalid, because the test relies on a form that the compiler does not offer.
We report validity as the proportion of generated tests that are valid.

\begin{table}[t]
\centering
\caption{Counts and causes of invalid tests summed over the three compilers, where P denotes \tool and A denotes \agentonly, and the last row gives the reduction from A to P.
\tool generates up to 97.00\% fewer invalid tests than \agentonly.}
\label{tab:rq1-causes}
\footnotesize
\let\approachfontsize\relax
\setlength{\tabcolsep}{3pt}
\renewcommand{\arraystretch}{1.12}
\begin{tabular}{@{}l>{\columncolor{toolbg}\centering\arraybackslash}p{2.3em}>{\centering\arraybackslash}p{2.3em}>{\columncolor{toolbg}\centering\arraybackslash}p{2.3em}>{\centering\arraybackslash}p{2.3em}>{\columncolor{toolbg}\centering\arraybackslash}p{2.3em}>{\centering\arraybackslash}p{2.3em}>{\columncolor{toolbg}\centering\arraybackslash}p{2.3em}>{\centering\arraybackslash}p{2.3em}>{\columncolor{toolbg}\centering\arraybackslash}p{2.3em}>{\centering\arraybackslash}p{2.3em}>{\columncolor{toolbg}\centering\arraybackslash}p{2.3em}>{\centering\arraybackslash}p{2.3em}@{}}
\toprule
\multirow{2}{*}{\textbf{Cause}} & \multicolumn{2}{c}{\textbf{GPT-5.5}} & \multicolumn{2}{c}{\textbf{GPT-6.1}} & \multicolumn{2}{c}{\textbf{Opus-5.5}} & \multicolumn{2}{c}{\textbf{Qwen-3.8}} & \multicolumn{2}{c}{\textbf{DeepSeek-4.1}} & \multicolumn{2}{c}{\textbf{MiniMax-3.1}} \\
\cmidrule(lr){2-3}\cmidrule(lr){4-5}\cmidrule(lr){6-7}\cmidrule(lr){8-9}\cmidrule(lr){10-11}\cmidrule(lr){12-13}
 & \multicolumn{1}{>{\columncolor{toolbg}}c}{P} & \multicolumn{1}{c}{A} & \multicolumn{1}{>{\columncolor{toolbg}}c}{P} & \multicolumn{1}{c}{A} & \multicolumn{1}{>{\columncolor{toolbg}}c}{P} & \multicolumn{1}{c}{A} & \multicolumn{1}{>{\columncolor{toolbg}}c}{P} & \multicolumn{1}{c}{A} & \multicolumn{1}{>{\columncolor{toolbg}}c}{P} & \multicolumn{1}{c}{A} & \multicolumn{1}{>{\columncolor{toolbg}}c}{P} & \multicolumn{1}{c}{A} \\
\midrule
\multicolumn{13}{@{}l}{\textit{Test construction errors}} \\
\quad API / Script Errors & 12 & 400 & 14 & 108 & 2 & 1 & 2 & 41 & 1 & 1 & 2 & 15 \\
\quad Malformed Graphs & 1 & 127 & 3 & 33 & \zero & 2 & 2 & 8 & 1 & 1 & \zero & 5 \\
\quad Missing Fixtures & \zero & 18 & \zero & 15 & \zero & \zero & \zero & \zero & \zero & \zero & \zero & \zero \\
\quad Precheck Failures & \zero & 1 & 5 & \zero & \zero & \zero & 3 & 2 & 2 & \zero & \zero & 1 \\
\quad Unsupported Operators & 3 & 183 & 46 & 3 & 10 & 7 & 1 & 4 & \zero & \zero & \zero & \zero \\
\multicolumn{13}{@{}l}{\textit{Oracle errors}} \\
\quad Incorrect Expected Outputs & 4 & \zero & 7 & \zero & 5 & \zero & \zero & 1 & \zero & \zero & \zero & \zero \\
\quad Inappropriate Numerical Tolerances & \zero & \zero & 1 & \zero & \zero & \zero & \zero & \zero & \zero & 1 & \zero & \zero \\
\quad Unverifiable Oracles & 2 & 5 & 3 & 2 & 11 & 2 & 1 & \zero & 1 & 2 & \zero & \zero \\
\midrule
\textbf{\# Invalid Tests} & \textbf{22} & \textbf{734} & \textbf{79} & \textbf{161} & \textbf{28} & \textbf{12} & \textbf{9} & \textbf{56} & \textbf{5} & \textbf{5} & \textbf{2} & \textbf{21} \\
\textit{Reduction} & \multicolumn{2}{c}{\textit{$-$97.0\%}} & \multicolumn{2}{c}{\textit{$-$50.9\%}} & \multicolumn{2}{c}{\textit{$+$133.3\%}} & \multicolumn{2}{c}{\textit{$-$83.9\%}} & \multicolumn{2}{c}{\textit{$+$0.0\%}} & \multicolumn{2}{c}{\textit{$-$90.5\%}} \\
\bottomrule
\end{tabular}

\end{table}

Figure~\ref{fig:rq1-validity} shows that 94.91\% of all tests generated by \tool are valid, against 72.09\% for \agentonly, and the validity of \tool with every agent is at least 89.89\%.
\tool improves the validity of \agentonly by at least 49.75 percentage points with GPT-5.5, Qwen-3.8, and MiniMax-3.1, and by up to 90.07 percentage points on TVM with GPT-5.5.
For example, \agentminimax fails to generate any test for TVM within 1,800 seconds per agent call, and only three tests on ONNX-MLIR and GeneSys are valid.
With the same agent, \propminimax reaches a validity of 96.00\%, 100.00\%, and 95.00\% on TVM, ONNX-MLIR, and GeneSys.
This is because \tool generates each test in short steps, where skeletons constrain each step and its validator and execution feedback catch the remaining errors.
For example, the validator rejects the first TVM candidate of \propminimax, which calls \codefont{relax.BlockBuilder.emit\_input} and \codefont{tvm.nd.array}, two functions that TVM 0.23.0 does not provide, and \propminimax then generates 24 valid tests on TVM.
In contrast, each call of \agentonly explores the compiler and writes a complete test script in one step, and no validator checks the script before it runs.

\agentonly reaches a higher validity than \tool in three campaigns, \ie on GeneSys with GPT-6.1 and DeepSeek-4.1 and on TVM with Opus-5.5, because its tests stay on forms that the compiler is known to accept.
For example, 91.32\% of the valid tests of \agentgptsix on GeneSys build a model around \codefont{Conv}, and 96.49\% only check output shapes and data types.
The scripts of \agentonly even state this choice in their comments.
One test of \agentdeepseek ``deliberately uses a conservative op set that GeneSys' ONNX front end is known to accept''.
Another test, by \agentopus on TVM, notes that TVM crashed with a segmentation fault when it used \codefont{R.shape\_of} to compute the target shape of \codefont{R.reshape}, so the test avoids this form.
In contrast, \tool tests 31, 35, and 110 operators in these campaigns, against 7, 8, and 15 of \agentonly.
These operators include harder cases that \agentonly avoids, and the invalid tests of \tool come from two kinds of them.
Some build legal forms that the compiler declares unsupported, such as adding a 1-D tensor to a 2-D tensor, which ONNX allows through broadcasting but GeneSys rejects.
Others use extreme inputs, on which the LLM computes a wrong expected output.
Giving the validator more compiler knowledge, such as the tensor ranks that each GeneSys template supports, would let \tool steer such candidates toward supported forms before execution.

\begin{table}[t]
\centering
\caption{Alarms raised by generated tests and their judgments.
An alarm is a crash, hang, exception, or failed assertion that a test reports when it runs on the compiler.
It is a bug when it comes from a compiler fault, and a false positive (FP) when the compiler is correct but the test is wrong, such as a wrong expected output or a form that the compiler declares unsupported.
It is numerical instability (NI) when the compiler uses a mathematically correct formula whose results fall outside the oracle's tolerance, while a more careful algorithm stays within it in the same data type.
The false-positive rate (FPR, \%) is the number of false positives divided by the number of alarms, and a dash marks a campaign without alarms.
The false-positive rate of \tool stays at or below 13.56\% in all but three campaigns with alarms.}
\label{tab:rq1-alarms}
\footnotesize
\let\approachfontsize\relax
\setlength{\tabcolsep}{2.6pt}
\renewcommand{\arraystretch}{1.15}
\begin{tabular}{@{}lrrrrrrrrrrrrrrr@{}}
\toprule
\multirow{2}{*}{\textbf{Approach}} & \multicolumn{5}{c}{\textbf{TVM}} & \multicolumn{5}{c}{\textbf{ONNX-MLIR}} & \multicolumn{5}{c}{\textbf{GeneSys}} \\
\cmidrule(lr){2-6}\cmidrule(lr){7-11}\cmidrule(l){12-16}
 & Alarm & Bug & FP & NI & FPR & Alarm & Bug & FP & NI & FPR & Alarm & Bug & FP & NI & FPR \\
\midrule
\propgptfive & 5 & 2 & 3 & \zero & 60.00 & \zero & \zero & \zero & \zero & -- & 48 & 45 & 3 & \zero & 6.25 \\
\agentgptfive & 24 & 23 & 1 & \zero & 4.17 & \zero & \zero & \zero & \zero & -- & 317 & 132 & 185 & \zero & 58.36 \\
\midrule
\propgptsix & 80 & 25 & 7 & 48 & 8.75 & 6 & 6 & \zero & \zero & \zero & 160 & 115 & 45 & \zero & 28.12 \\
\agentgptsix & \zero & \zero & \zero & \zero & -- & \zero & \zero & \zero & \zero & -- & 104 & 101 & 3 & \zero & 2.88 \\
\midrule
\propopus & 45 & 30 & 4 & 11 & 8.89 & 28 & 26 & 2 & \zero & 7.14 & 59 & 51 & 8 & \zero & 13.56 \\
\agentopus & \zero & \zero & \zero & \zero & -- & \zero & \zero & \zero & \zero & -- & 42 & 35 & 7 & \zero & 16.67 \\
\midrule
\propqwen & 1 & 1 & \zero & \zero & \zero & \zero & \zero & \zero & \zero & -- & 5 & 4 & 1 & \zero & 20.00 \\
\agentqwen & \zero & \zero & \zero & \zero & -- & \zero & \zero & \zero & \zero & -- & 11 & 7 & 4 & \zero & 36.36 \\
\midrule
\propdeepseek & 3 & \zero & \zero & 3 & \zero & \zero & \zero & \zero & \zero & -- & 12 & 12 & \zero & \zero & \zero \\
\agentdeepseek & \zero & \zero & \zero & \zero & -- & \zero & \zero & \zero & \zero & -- & \zero & \zero & \zero & \zero & -- \\
\midrule
\propminimax & \zero & \zero & \zero & \zero & -- & \zero & \zero & \zero & \zero & -- & 5 & 5 & \zero & \zero & \zero \\
\agentminimax & \zero & \zero & \zero & \zero & -- & \zero & \zero & \zero & \zero & -- & 1 & 1 & \zero & \zero & \zero \\
\bottomrule
\end{tabular}

\end{table}

To see why tests become invalid, we break the invalid tests of each approach down by cause, as reported in Table~\ref{tab:rq1-causes}.
\tool generates 85.34\% fewer invalid tests than \agentonly overall and 97.00\% fewer with GPT-5.5.
This is because \agentonly repeats the same mistakes across iterations.
For example, 232 tests of \agentgptfive on TVM fail with the same error because \codefont{tvm.nd} no longer exists, and the error also appears in 82 tests of \agentgptsix.
Execution feedback in \tool reduces this error to two tests for \propgptfive, and its validity on TVM rises from 91.20\% in the first three hours of the campaign to 97.90\% in the last three hours.
In contrast, \agentgptfive does not receive such feedback, and its validity on TVM stays low, at 7.00\% and 2.60\% in the same periods.
Similarly, on GeneSys, \agentgptfive generates 183 models with forms that GeneSys declares unsupported, while \propgptfive generates only three.
The open-weight agents repeat their mistakes in the same way.
For example, 36.11\% of the invalid tests of \agentqwen on GeneSys create a random number generator with Python's \codefont{random.Random} and then call a method that only NumPy's generator provides, such as \codefont{integers}, so the script raises an \codefont{AttributeError} before it builds a model.
Validation of \tool also stops such errors before execution, and it rejects 6.73\% of the candidates of \propgptfive.
For example, it rejects four TVM candidates of \propqwen that write the operator \codefont{!=} as \codefont{\textbackslash!=}, which Python cannot parse, and a \codefont{ReduceMean} candidate of \propgptfive on ONNX-MLIR that passes the dimensions to average over as a second input tensor, which this operator requires in its \codefont{axes} setting.

\begin{figure*}[t]
\centering
\input{figures/rq2-diversity-fig-v1}
\caption{Diversity: cumulative counts of distinct operator--property--generator combinations, operators, properties, and data generators in valid tests over each six-hour campaign, for each compiler and agent.
Each panel compares \tool with \agentonly under the same agent, the shading marks the difference between them, and the numbers give the final counts.
A combination counts once, however many valid tests exercise it.
\tool exercises more combinations than \agentonly in 14 of the 18 campaigns, and more operators, properties, and data generators in 17 each.}
\label{fig:rq2-diversity}
\end{figure*}

An incorrect oracle can also make a correct compiler look faulty, so we next examine how many alarms of the generated tests are false positives, as Table~\ref{tab:rq1-alarms} shows.
The false-positive rate of \tool stays at or below 13.56\% except in three campaigns, whose false positives are unsupported forms or incorrect expected outputs.
On GeneSys, every false positive of \propgptsix and \propqwen is a form that GeneSys declares unsupported.
The false positives of \tool on TVM come from incorrect expected outputs.
For example, TVM's \codefont{relax.median} returns both the median values and their positions, but a test of \propgptsix expects a single tensor.

Another 62 tests generated by \tool on TVM expose numerical instability, where TVM uses a mathematically correct formula whose results fall outside the oracle's tolerance, while a more careful algorithm stays within it in the same data type.
For example, TVM computes \codefont{sinh} as \codefont{(exp(x)-exp(-x))/2}, which is mathematically correct but loses precision for inputs close to zero, where the two exponentials nearly cancel, and overflows for inputs at which \codefont{sinh} is still finite.
A test of \propgptsix therefore sees a larger error than its oracle allows, while NumPy stays within the same tolerance in \codefont{float32}.
Similarly, TVM computes \codefont{atan(x)} as \codefont{asin(x / sqrt(x*x + 1))}, so a test of \propdeepseek that passes infinite inputs receives NaN where the expected output is $\pm\pi/2$.
We count these tests as valid and report them separately from compiler faults.
\begin{rqsummary}{1}
\tool reaches a validity of 94.91\% across all its generated tests, against 72.09\% for \agentonly, and generates 85.34\% fewer invalid tests.
With GPT-5.5, Qwen-3.8, and MiniMax-3.1, \tool improves validity by at least 49.75 percentage points, because skeletons constrain each generation step and validation and feedback catch the remaining errors.
\end{rqsummary}

\subsection{RQ2: Test Diversity}
\label{sec:rq2}
Test diversity demonstrates the extent to which a set of tests can explore the semantic space of a compiler.
To assess it, we compare \tool and \agentonly under the same configuration by the distinct operators, properties, data generators, and operator--property--generator combinations used by their valid tests.
For \tool, an operator counts when a test's property check targets it, which we determine from the test's assertions.
For \agentonly, we count each operator that an assertion checks, or every operator of the graph when the assertion checks the output of the whole graph.
Each counted operator forms a combination with the test's property and data generator, and repeated valid tests of the same combination count once.
We map \agentonly's labels to \tool's vocabulary where possible and retain unmatched property labels.

\begin{figure*}[t]
\centering
\input{figures/eval-style}
\pgfplotsset{table/col sep=comma}
\def\evalcellw{0.793pt}\def\evalcellh{5.1pt}
\begin{tikzpicture}
\begin{groupplot}[group style={group name=cov, group size=3 by 2, horizontal sep=8pt, vertical sep=7pt}, eval coverage, height=95.19999999999999pt, ymin=0.5, ymax=17.5, ytick={1,...,17}]
\nextgroupplot[width=138.75pt, xmin=0.5, xmax=175.5, yticklabels={{Reference Consistency},{Shape and Dtype Contract},{Cancellation},{Commutativity},{Identity},{Concrete Input Replay},{Geometric Decomposition},{Monotonicity},{Shift Invariance},{Decomposition Identity},{Idempotence},{Positive Homogeneity},{Distributivity},{Permutation Invariance},{Associativity},{Mathematical Semantics Consistency},{Combined Geometric Decomposition}}, ylabel={\toolname{}}, title={TVM}]
\addplot[eval cell=evalPropilot, keep panel=1] table[x=operator_index, y=property_index] {figures/data/rq2-coverage.csv};
\nextgroupplot[width=131.61pt, xmin=0.5, xmax=166.5, yticklabels={}, title={ONNX-MLIR}]
\addplot[eval cell=evalPropilot, keep panel=3] table[x=operator_index, y=property_index] {figures/data/rq2-coverage.csv};
\nextgroupplot[width=39.64pt, xmin=0.5, xmax=50.5, yticklabels={}, title={GeneSys}]
\addplot[eval cell=evalPropilot, keep panel=5] table[x=operator_index, y=property_index] {figures/data/rq2-coverage.csv};
\nextgroupplot[width=138.75pt, xmin=0.5, xmax=175.5, yticklabels={{Reference Consistency},{Shape and Dtype Contract},{Cancellation},{Commutativity},{Identity},{Concrete Input Replay},{Geometric Decomposition},{Monotonicity},{Shift Invariance},{Decomposition Identity},{Idempotence},{Positive Homogeneity},{Distributivity},{Permutation Invariance},{Associativity},{Mathematical Semantics Consistency},{Combined Geometric Decomposition}}, ylabel={\agentname{}}, xlabel={175 Operators}]
\addplot[eval cell=evalAgent, keep panel=2] table[x=operator_index, y=property_index] {figures/data/rq2-coverage.csv};
\nextgroupplot[width=131.61pt, xmin=0.5, xmax=166.5, yticklabels={}, xlabel={166 Operators}]
\addplot[eval cell=evalAgent, keep panel=4] table[x=operator_index, y=property_index] {figures/data/rq2-coverage.csv};
\nextgroupplot[width=39.64pt, xmin=0.5, xmax=50.5, yticklabels={}, xlabel={50 Operators}]
\addplot[eval cell=evalAgent, keep panel=6] table[x=operator_index, y=property_index] {figures/data/rq2-coverage.csv};
\end{groupplot}
\end{tikzpicture}
\caption{Operator--property pairs exercised by the valid tests of all six agents combined, for each compiler.
Rows are properties, columns are operators, and a filled cell marks a pair that at least one valid test exercises.
The last row, Combined Geometric Decomposition, counts tests that check geometric decomposition and another property in one assertion, such as \codefont{op(x, y) == concat(op(y1, x1), op(y2, x2))} for commutativity (Section~\ref{sec:property-skeletons}).
\tool covers at least 2.73$\times$ the pairs of \agentonly on each compiler, and on TVM, six of the 17 property rows are empty for \agentonly.}
\label{fig:rq2-coverage}
\end{figure*}

Figure~\ref{fig:rq2-diversity} plots how these four counts grow over each six-hour campaign.
Overall, \tool covers more operator--property--generator combinations than \agentonly in 14 of the 18 campaigns, and more operators, properties, and data generators in 17 campaigns each.
With the three proprietary agents, \tool exercises on average 3.59$\times$ the combinations of \agentonly, and up to 9.50$\times$ on TVM with GPT-5.5.
All three parts of a combination contribute to this gain, as \tool tests on average 4.67$\times$, 4.48$\times$, and 4.97$\times$ the operators, properties, and data generators of \agentonly with these agents, respectively.
For example, \propgptsix tests 171 operators on TVM, including \codefont{eye\_like} and \codefont{meshgrid}, but \agentgptsix never invokes 142 of them.
This is because \tool records the number of tests that have used each operator and feeds these usage records back to the agent, which steers it toward untested operators, so \propgptsix spends at most four valid tests on any TVM operator.
In contrast, \agentgptsix does not have such usage records as feedback, and 94.05\% of its valid tests target \codefont{add}.
This advantage does not depend on the characteristics of a particular agent.
With each of GPT-5.5, GPT-6.1, and Opus-5.5, \tool tests more operators than \agentonly on all three compilers, with an average gain of at least 4.50$\times$ per agent.
With the open-weight agents, \tool tests more operators in eight of the nine campaigns, such as 51 TVM operators of \propdeepseek against 16 of \agentdeepseek, and it uses more properties and more data generators in all nine.

\agentonly exercises more combinations in four campaigns.
On GeneSys with GPT-6.1, \propgptsix only checks that GeneSys compiles each model and runs it without error.
This is because the validator rejected the first candidate of this campaign, a check of output shapes and data types, and the agent then chose the compile-and-run check for all later tests.
Nevertheless, we still observe that \propopus uses 16 properties on GeneSys, such as distributivity and monotonicity, and verifies them on the reference, so a validator that accepts every property the reference can verify would keep such checks.
The other three campaigns use open-weight agents, namely Qwen-3.8 on ONNX-MLIR and GeneSys and DeepSeek-4.1 on ONNX-MLIR.
In these campaigns, each valid test of \agentonly contributes more than three combinations on average, because it checks the output of a whole graph and every operator of that graph counts, while each test of \tool targets one operator.
Nevertheless, we still observe that \propqwen uses 16 properties and 19 data generators on ONNX-MLIR, while \agentqwen uses three properties and one data generator only.

To better understand whether the tests spread their checks across the compiler or concentrate on a few operators and properties, we also profile the operator--property coverage of valid tests combining the six agents, as shown in Figure~\ref{fig:rq2-coverage}.
Overall, \tool covers at least 2.73$\times$ the operator--property pairs of \agentonly on every compiler.
We observe that the pairs of \agentonly cluster on common operators, while \tool extends its pairs to operators that \agentonly never tests.
For example, on TVM, 25.21\% of \agentonly's pairs fall on \codefont{permute\_dims}, \codefont{add}, and \codefont{sum}, while 75.25\% of the pairs of \tool lie on operators that \agentonly never tests.
This is because usage records steer the agent toward untested operators, and the applicability conditions of the skeletons let it pair each new operator with the properties that apply to it.
On TVM, for instance, \tool pairs each operator with a median of three properties.
In contrast, 92.14\% of the valid tests of \agentopus on GeneSys check \codefont{Gemm}.

\begin{table}[t]
  \centering
  \caption{Distinct symptoms detected by each approach, by kind and by the oracle that first detects each symptom.
  Two failing tests show the same symptom when they fail with the same error message at the same code location or, for silent wrong outputs, when the same operator returns wrong outputs.
  The implicit oracle flags crashes, hangs, and exceptions without any assertion.
  Each row counts the symptoms of one agent, so a symptom found with several agents appears in several rows.
  Combined, \tool detects 81 distinct symptoms, while \agentonly detects 13.
  Only \tool observes silent wrong outputs, which its algebraic relations, shape checks, and reference comparisons detect, whereas every symptom of \agentonly is detected by the implicit oracle.}
  \label{tab:rq3-tools}
  \footnotesize
\let\approachfontsize\relax
\setlength{\tabcolsep}{3pt}
\renewcommand{\arraystretch}{1.15}
\begin{tabular}{@{}lccccc@{\hspace{8pt}}cccc@{}}
\toprule
\multirow{2}{*}{\textbf{Approach}} & \multicolumn{5}{c}{\textbf{\# Symptoms}} & \multicolumn{4}{c}{\textbf{\# Symptoms by Oracle}} \\
\cmidrule(lr){2-6}\cmidrule(l){7-10}
 & \textbf{Crash} & \textbf{Hang} & \textbf{Exception} & \makecell[b]{\textbf{Silent}\\\textbf{Wrong Output}} & \textbf{Total} & \makecell[b]{\textbf{Algebraic}\\\textbf{Relation}} & \makecell[b]{\textbf{Shape}\\\textbf{Check}} & \makecell[b]{\textbf{Reference}\\\textbf{Comparison}} & \makecell[b]{\textbf{Implicit}\\\textbf{Oracle}} \\
\midrule
\propgptfive & \zero & \zero & \textbf{11} & \zero & \textbf{11} & \zero & \zero & \zero & \textbf{11} \\
\agentgptfive & \textbf{1} & \zero & 6 & \zero & 7 & \zero & \zero & \zero & 7 \\
\midrule
\propgptsix & \zero & \zero & \textbf{23} & \textbf{10} & \textbf{33} & \textbf{3} & \textbf{1} & \textbf{6} & \textbf{23} \\
\agentgptsix & \zero & \zero & 5 & \zero & 5 & \zero & \zero & \zero & 5 \\
\midrule
\propopus & \textbf{3} & \textbf{1} & \textbf{35} & \textbf{23} & \textbf{62} & \textbf{15} & \zero & \textbf{8} & \textbf{39} \\
\agentopus & \zero & \zero & 4 & \zero & 4 & \zero & \zero & \zero & 4 \\
\midrule
\propqwen & \zero & \zero & 3 & \zero & 3 & \zero & \zero & \zero & 3 \\
\agentqwen & \zero & \zero & \textbf{5} & \zero & \textbf{5} & \zero & \zero & \zero & \textbf{5} \\
\midrule
\propdeepseek & \zero & \zero & \textbf{7} & \zero & \textbf{7} & \zero & \zero & \zero & \textbf{7} \\
\agentdeepseek & \zero & \zero & \zero & \zero & \zero & \zero & \zero & \zero & \zero \\
\midrule
\propminimax & \zero & \zero & \textbf{2} & \zero & \textbf{2} & \zero & \zero & \zero & \textbf{2} \\
\agentminimax & \zero & \zero & 1 & \zero & 1 & \zero & \zero & \zero & 1 \\
\bottomrule
\end{tabular}

\end{table}

Within each property row of Figure~\ref{fig:rq2-coverage}, \tool also pairs the property with more operators.
For example, on TVM, \tool pairs monotonicity with 46 operators and geometric decomposition with 64, but both rows are empty for \agentonly.
This is because each skeleton lists the operator classes that its property applies to, such as order-preserving elementwise operators for monotonicity, so the agent can apply one property to every operator in these classes.
Combining the six agents, six of the 17 property rows on TVM are empty for \agentonly, which means that no test of \agentonly checks, for example, whether regrouping additions preserves the result or whether a larger input to an order-preserving operator gives a larger output.
Even on the operators that both approaches test, \tool checks each operator with more properties on average on every compiler.
For example, on ONNX-MLIR, \tool checks \codefont{Conv} with 11 properties, including monotonicity and positive homogeneity, but \agentonly only checks it with five.
\begin{rqsummary}{2}
\tool exercises more operator--property--generator combinations than \agentonly in 14 of the 18 campaigns, on average 3.59$\times$ and up to 9.50$\times$ with the proprietary agents, and it tests more operators, properties, and data generators, each in 17 campaigns.
Usage records steer the agent toward untested operators, and skeletons provide properties that \agentonly does not use.
\end{rqsummary}

\begin{figure*}[t]
\centering
\input{figures/rq3-bug-map-fig-v1}
\caption{Distinct root causes detected by the six agents combined, grouped by the defect class in the compiler (rows) and by compiler and approach (columns).
Each square is one root cause, and its shading gives the kind of symptom that the root cause produces.
\tool detects 96 root causes across all nine defect classes, including 25 that produce silent wrong outputs, whereas \agentonly detects 13, 11 of which make the compiler raise an error during compilation.}
\label{fig:rq3-bug-map}
\end{figure*}

\subsection{RQ3: Test Usefulness}
\label{sec:rq3}
We next evaluate whether the generated tests detect faults in the three target compilers.
A failing test shows one of four kinds of symptoms: the compiler crashes, hangs, or raises an exception, or the compiled model runs without any error but returns a wrong result, which we call a silent wrong output.
Two failing tests show the same symptom when they fail in the same way with the same error message raised at the same code location, or, for silent wrong outputs, when the same operator returns wrong outputs.
We then analyze the failing tests of each symptom, locate the faults in the compiler source code, and count each root cause once.
Explicit rejections of unsupported operators and numerical deviations of mathematically correct formulas do not count as faults.

\MyPara{Symptoms.}
We start from what the failing tests observe, and Table~\ref{tab:rq3-tools} counts the distinct symptoms of each approach and the oracle that first detects each of them.
Combining the six agents, tests generated by \tool show 81 distinct symptoms, 6.23$\times$ those of \agentonly, and \tool shows more symptoms with five of the six agents.
Silent wrong outputs appear only in the tests of \tool, all of them with \propgptsix and \propopus.

Algebraic relations detect three silent wrong outputs of \propgptsix and 15 of \propopus, and these outputs are wrong in different ways.
For example, \codefont{not\_equal} should report a value as unequal to itself exactly when the value is NaN (\ie\ \codefont{not\_equal(x, x) == isnan(x)}), because IEEE~754 treats NaN as unequal to every value, including itself.
When \propopus checks this relation on TVM with inputs that contain NaN, the expected output is true at every NaN position, but the compiled model returns false.
Similarly, \codefont{adaptive\_avg\_pool2d} averages each channel separately, so pooling a tensor should give the same result as pooling its channels in two parts and concatenating the results (\ie\ \codefont{pool(x) == concat(pool(x[:, :2]), pool(x[:, 2:]))}).
When \propgptsix splits a tensor of five channels into two and three channels, the two results should be identical, but they differ in three output values, such as 6.17 for the whole tensor and 3.92 for its parts.
Likewise, \codefont{argmax} returns the position of the largest value, so adding the same constant to every element should not change that position (\ie\ \codefont{argmax(x + c) == argmax(x)}).
\propopus checks this relation on a corner case where every element equals the lowest value of its data type, such as 0 for \codefont{uint8}, so all elements tie and the expected position is the first one, 0.
The compiled model returns $-1$, which is not a valid position, although it returns 0 once the constant is added.
This is because each skeleton states a relation that every valid input must satisfy, and the data generators place inputs at corner cases where compiled code tends to diverge, such as NaN from Semantic Gap Witnesses, channel splits from Axis Split, and the lowest value of a data type from Data Type Boundaries.
In contrast, every symptom of \agentonly is detected by the implicit oracle, which flags crashes, hangs, and exceptions without any assertion, and none of them is detected by an assertion that \agentonly wrote.

Exceptions are the most frequent kind of symptom for both approaches, and they are the only kind that \tool shows with the open-weight agents.
An exception counts as a symptom of a fault only when it is not an expected rejection, that is, when the compiler accepts the model instead of reporting the operator or its form as unsupported.
For example, \propgptfive checks that a softmax over the last axis of a $1\times4$ tensor keeps the shape and data type of its input.
GeneSys has a code template for two-dimensional softmax, so it imports the model and selects this template, but it then stops with \codefont{KeyError: `SHIFTR'} because the template uses an instruction that GeneSys does not define.
Similarly, \codefont{Clip} limits each value to a range, so it preserves the order of two inputs (\ie\ \codefont{x <= y} implies \codefont{Clip(x) <= Clip(y)}).
\propdeepseek checks this relation on GeneSys with a model under ONNX opset 6, which gives the range of \codefont{Clip} as two attributes of the node.
The ONNX checker accepts the model, and GeneSys does not report the opset as unsupported, but its importer stops with \codefont{AttributeError: `NoneType' object has no attribute `graph'}.
In both cases, the compiler accepts the input and then fails in its own code instead of reporting the input as unsupported.
This is because the data generators produce legal but uncommon forms, such as the single-row tensor from Shape Edge Cases, and execution feedback leads the agent to another legal form when a model is rejected.
For example, \propdeepseek turns to opset 6 after the ONNX checker rejects its previous \codefont{Clip} model, which gives the range as attributes under opset 11, where \codefont{Clip} takes it as inputs.
In contrast, \agentonly shows at most six exception symptoms with any agent.

\begingroup
\newcommand{\examplefont}{\fontfamily{lmtt}\fontsize{7}{8.5}\selectfont}
\lstdefinestyle{rqexample}{
basicstyle=\examplefont\color{black},
lineskip=0.6pt,
columns=fullflexible,
keepspaces=true,
numbers=none,
frame=tb,
framerule=0.35pt,
rulecolor=\color{black!40},
xleftmargin=0pt,
framexleftmargin=0pt,
showstringspaces=false,
keywordstyle=\bfseries\color{black},
identifierstyle=\color{black},
stringstyle=\color{black},
commentstyle=\color[HTML]{1A4A8A},
escapeinside={(*}{*)}
}
\begin{figure}[t]
\begin{minipage}[t]{0.53\linewidth}
\begin{lstlisting}[style=rqexample,language=Python]
# Property: cancellation, reshape-compatible input
# Two channels with distinct values
x = rng.uniform(-8, 8, (1, 2, 4, 6)).astype(np.float32)
nodes = [
    make_node("SpaceToDepth", ["x"], ["Y"], blocksize=2),
    make_node("DepthToSpace", ["Y"], ["Z"], blocksize=2,
              mode="DCR")]
# Expected outputs follow the ONNX definition.
y = space_to_depth(x, 2)    # perm: (0, 3, 5, 1, 2, 4)
z = depth_to_space_dcr(y, 2)
assert np.array_equal(z, x)
\end{lstlisting}
\end{minipage}\hfill
\begin{minipage}[t]{0.44\linewidth}
\begin{lstlisting}[style=rqexample,language=C++]
// SpaceToDepth.cpp in ONNX-MLIR
// Reshape input to [B, C, H/bs, bs, W/bs, bs].
Value reshapeRes1 =
    create.krnlOnnx.reshape(input, outputDims1);
// Transpose into [B, C, bs, bs, H/bs, W/bs].
(*{\setlength{\fboxsep}{0.6pt}\colorbox[HTML]{F6C9CD}{SmallVector<int64\_t> perm(\{0, 1, 3, 5, 2, 4\});}}*)
// ONNX requires perm({0, 3, 5, 1, 2, 4}),
// i.e., [B, bs, bs, C, H/bs, W/bs].
Value transposeRes = create.krnlOnnx
    .transpose(reshapeRes1, perm, outputDims2);
\end{lstlisting}
\end{minipage}
\captionof{lstlisting}{A cancellation test generated by \propopus (left) and the \codefont{SpaceToDepth} lowering of ONNX-MLIR that it exposes (right), where the highlighted permutation places the input channel before the block positions, and the comment that we add below it gives the permutation that ONNX defines.}
\label{lst:rq3-spacetodepth}
\end{figure}
\endgroup

\propopus also exposes three crashes during compilation by ONNX-MLIR.
For example, adding a constant to the input should not change the output of \codefont{MeanVarianceNormalization}, which subtracts the mean and divides by the standard deviation along the given axes (\ie\ \codefont{MVN(x + c) == MVN(x)}).
When \propopus names the last axis as $-1$ (\ie\ \codefont{axes=[-1]}), the expected output is the normalized tensor, but ONNX-MLIR aborts on an internal assertion that an index is within bounds.
Failure-adjacent exploration then directs the agent to a related operator, and \propopus applies the same negative axis to \codefont{LpNormalization}, which crashes ONNX-MLIR on a similar assertion.
The third crash comes from \codefont{LRN}, which divides each value by a term computed from its neighboring channels and returns its input unchanged when \codefont{alpha=0} and \codefont{bias=1} (\ie\ \codefont{LRN(x, alpha=0, bias=1) == x}).
On a \codefont{float64} input, ONNX-MLIR aborts on an assertion that both operands of a multiplication have the same type.
Identity Injection chooses such parameters, so the test knows its expected output without a reference implementation.
In contrast, the only crash of \agentonly is a segmentation fault of TVM when \agentgptfive declares a tensor shape as \codefont{R.Tensor((None,))}.

The hang also comes from an identity test of \propopus.
\codefont{Shrink} sets values near zero to zero and shifts the others toward zero, and it returns its input unchanged when \codefont{lambd=0} and \codefont{bias=0} (\ie\ \codefont{Shrink(x, lambd=0, bias=0) == x}).
The model holds a single \codefont{Shrink} node on a $2\times3$ \codefont{int32} tensor, and ONNX-MLIR compiles the models of this campaign in a median of 2.46 seconds, but the compilation of this model reaches a code path marked unreachable and does not finish within 300 seconds.
The agent chose the \codefont{int32} input because failure-adjacent exploration directed it toward integer data types after earlier integer tests had failed, such as a \codefont{Less} model on \codefont{uint32} inputs.

\MyPara{Root Causes.}
Analyzing the failing tests of these symptoms yields the root causes in Figure~\ref{fig:rq3-bug-map}, grouped by the defect in the compiler.
Combining the six agents, \tool detects 96 distinct root causes, 7.38$\times$ the 13 detected by \agentonly, and it detects more root causes with five of the six agents, on average 9.39$\times$ with the proprietary agents and up to 18.75$\times$ with Opus-5.5, where \propopus detects 75 and \agentopus detects four.
\tool covers all nine defect classes, including 25 root causes that produce silent wrong outputs, while 11 of the 13 root causes of \agentonly make the compiler raise an error during compilation, such as GeneSys's importer rejecting the standard \codefont{count\_include\_pad} attribute of \codefont{AveragePool}.

For example, \propopus detects an incorrect operator lowering in ONNX-MLIR, shown in Listing~\ref{lst:rq3-spacetodepth}.
The operator \codefont{SpaceToDepth} cuts each channel into $2\times2$ blocks and turns the four values of each block into four channels, and \codefont{DepthToSpace} reverses this, so applying the two in turn should restore the input (\ie\ \codefont{DepthToSpace(SpaceToDepth(x)) == x}).
ONNX orders the new channels by the position in the block first and by the input channel second.
For a $2\times2$ input whose two channels hold $a,b,c,d$ and $e,f,g,h$, \codefont{SpaceToDepth} should return the channels $a,e,b,f,c,g,d,h$.
The lowering in ONNX-MLIR instead transposes with the highlighted permutation \codefont{\{0, 1, 3, 5, 2, 4\}}, which keeps the input channel first and returns $a,b,c,d,e,f,g,h$, while ONNX requires \codefont{\{0, 3, 5, 1, 2, 4\}}.
The two orders agree only when the input has a single channel, so the test of \propopus, which uses two channels with distinct values, receives the values in a different order after \codefont{DepthToSpace}.
Similarly, TVM computes \codefont{asin} of an input below 0.5 with a Taylor series, where each term multiplies the previous one by $x^2$ and a fixed ratio, and the ratio for the third term should be 9/20.
TVM instead writes \codefont{term5 = term3 * x2 * 9 / 40}, so a reference comparison of \propgptsix finds outputs that differ from NumPy.
\tool also detects root causes with open-weight agents.
For example, the \codefont{Clip} test of \propdeepseek exposes an incorrect model import in GeneSys.
GeneSys's importer replaces the model with the result of shape inference and then reads its graph (\ie\ \codefont{onnx\_proto.graph.node}) without a check, although shape inference returns \codefont{None} for models below opset 7.

\begin{table*}[t]
  \centering
  \caption{Categories of the 161 faulty TVM versions in issue replication.}
  \label{tab:rq3-case-categories}
    \footnotesize
  \setlength{\tabcolsep}{4pt}
  \renewcommand{\arraystretch}{1.05}
  \begin{tabular}{@{}>{\raggedright\arraybackslash}p{0.14\textwidth}>{\raggedleft\arraybackslash}p{0.07\textwidth}>{\raggedright\arraybackslash}p{0.23\textwidth}>{\raggedright\arraybackslash}p{\dimexpr0.56\textwidth-6\tabcolsep\relax}@{}}
    \toprule
    \textbf{Category} & \textbf{\# Cases} & \textbf{Description} & \textbf{Examples} \\
    \midrule
    Operator Logic & 58 & Incorrect functions, operand order, comparisons, or constants &
    Maximum returns the smaller operand instead of the larger one.\newline
    Subtraction computes $b-a$ instead of $a-b$. \\
    \addlinespace[3pt]
    Floating-Point & 59 & Incorrect rounding, numerical constants, or precision &
    Addition computes $a(1+\epsilon)+b$ instead of $a+b$ for small $\epsilon$.\newline
    Addition drops four mantissa bits from the left operand. \\
    \addlinespace[3pt]
    Optimization and Lowering & 44 & Incorrect shape inference, simplification, or lowering &
    Broadcasting keeps a size-one dimension instead of matching the other input.\newline
    Floor division rounds negative quotients toward zero instead of down. \\
    \midrule
    \textbf{Total} & \textbf{161} & & \\
    \bottomrule
  \end{tabular}

\end{table*}

\begin{figure}[t]
\centering
\input{figures/eval-style}
\pgfplotsset{table/col sep=comma}
\begin{tikzpicture}
\begin{axis}[
  eval bars, scale only axis, width=260pt, height=0.95in,
  bar width=9pt, ymin=0, ymax=55, ytick={0,10,...,50},
  xmin=0.5, xmax=3.5, xtick={1,2,3}, restrict x to domain=1:3,
  xticklabels={Operator Logic, Floating-Point, Optimization / Lowering},
  ylabel={\# Detected Cases},
  nodes near coords={\pgfmathprintnumber[fixed,precision=0]{\pgfplotspointmeta}},
  legend columns=1, legend style={at={(1.03,0.5)}, anchor=west},
]
\addplot[eval bar=evalPropilot, eval texture=evalNE, bar shift=-18.75pt] table[x=x, y=propilot] {figures/data/rq3-replication.csv};
\addplot[eval bar=evalCrash, eval texture=evalNE, bar shift=-6.25pt] table[x=x, y=crash] {figures/data/rq3-replication.csv};
\addplot[eval bar=evalAgent, eval texture=evalNW, bar shift=6.25pt] table[x=x, y=agent] {figures/data/rq3-replication.csv};
\addplot[eval bar=evalRepaired, eval texture=evalNW, bar shift=18.75pt] table[x=x, y=repaired] {figures/data/rq3-replication.csv};
\legend{\toolname, \toolname{} (Crash Only), \agentname, \agentname{} (API Repaired)}
\end{axis}
\end{tikzpicture}
\caption{Fault detection by case category.
\propgptfive detects 2.63$\times$ the faulty versions that \agentgptfive detects, and crash detection alone finds five of its 100 detected faulty versions.}
\label{fig:rq3-replication}
\end{figure}

\agentonly misses these root causes because its tests never build the inputs that reach them.
\agentopus never invokes \codefont{SpaceToDepth} in its tests on ONNX-MLIR, \agentgptsix never invokes \codefont{asin} on TVM, where 94.05\% of its valid tests target \codefont{add}, and all GeneSys tests of \agentdeepseek use opset 13.
In contrast, usage records steer \tool toward operators that it has not yet tested, as shown in RQ2 (Section~\ref{sec:rq2}), so 90 of the 125 root causes that the three proprietary agents detect with \tool, counted per agent, lie on operators that \agentonly with the same agent does not test.

\agentonly detects more root causes in only two campaigns, on TVM with GPT-5.5 (three against one) and on GeneSys with Qwen-3.8 (five against two).
All three root causes of \agentgptfive come from invalid tests, while \propgptfive first checks only output shapes and data types on 156 TVM operators, and \propqwen generates only 19 tests on GeneSys, as it spends a median of 358 seconds generating and 301 seconds validating each candidate.
Skeletons that ask for semantic relations more explicitly, and validation that reviews candidates selectively or with a faster LLM, would close these gaps.

\MyPara{Issue Replication.}
We also evaluate whether generated tests detect 161 existing faulty versions of TVM, each with one fault in operator logic, floating-point behavior, or optimization and lowering, as listed in Table~\ref{tab:rq3-case-categories}.
We compare \propgptfive and \agentgptfive in this experiment.
A test detects a faulty version when it passes on the fixed TVM but reports a crash or a failed assertion on the faulty version.
We also evaluate \propgptfive with crash detection alone to measure the contribution of its semantic checks.
For \agentgptfive, we additionally replace its outdated \codefont{tvm.nd.array} calls with \codefont{tvm.runtime.tensor} and rerun the tests.

Figure~\ref{fig:rq3-replication} shows that \propgptfive detects 100 of the 161 faulty versions, 2.63$\times$ the 38 detected by \agentgptfive, and it detects more faulty versions in every category, such as 46 against 11 for operator logic.
Part of this gap comes from the outdated \codefont{tvm.nd.array} calls of \agentgptfive, which make its tests fail before they check the compiler.
With these calls repaired, \agentgptfive detects 53 faulty versions, and \propgptfive still detects 1.89$\times$ as many.

This is because the semantic checks of \propgptfive, rather than crashes, detect most faulty versions.
Of the 100 faulty versions that \propgptfive detects, only five make a test crash, and the other 95 are detected only because an assertion of the test fails.
Comparing compiled outputs with reference outputs detects 85 of these 100 faulty versions.
Another 11 are detected only by checks of the output shapes that TVM infers, because their faults change shape inference or constant folding but leave the numerical outputs of the default compilation pipeline unchanged.
For example, broadcasting should expand an input dimension of size one to match the other input, but one faulty version still gives this dimension size one in the inferred output shape.
An algebraic check also detects an addition fault that reference comparisons miss, where the faulty version scales $a$ before adding $b$ instead of computing $a+b$.
\begin{rqsummary}{3}
With five of the six agents, tests generated by \tool detect more root causes than \agentonly, on average 9.39$\times$ and up to 18.75$\times$ with the proprietary agents.
Combining the six agents, \tool detects 96 distinct root causes, including 25 with silent wrong outputs that \agentonly does not detect.
In issue replication, \tool detects 2.63$\times$ the faulty TVM versions that \agentonly detects.
\end{rqsummary}

\begin{table}[t]
  \centering
  \caption{Ablation on ONNX-MLIR with MiniMax-3.1, where each row adds one component to the row above: property skeletons, then validation, then feedback, which completes \propminimax.
  Validity is the proportion of valid tests, and the remaining columns count the distinct operators, properties, data generators, and operator--property--generator combinations in valid tests.
  Each configuration runs for six hours, and the best value in each column is in bold.
  Skeletons raise validity from 18.18\% to 93.33\%, validation raises it to 100.00\%, and feedback yields 1.47$\times$ the combinations.}
  \label{tab:rq4-ablation}
  \footnotesize
\let\approachfontsize\relax
\setlength{\tabcolsep}{5.5pt}
\renewcommand{\arraystretch}{1.15}
\begin{tabular}{@{}lccccc@{}}
\toprule
\textbf{Approach} & \textbf{Validity} & \textbf{\# Operators} & \textbf{\# Properties} & \textbf{\# Data Generators} & \textbf{\# Combinations} \\
\midrule
\agentminimax & 18.18\% & 15 & 2 & 1 & 16 \\
\plusskeleton & 93.33\% & 12 & 9 & 10 & 14 \\
\plusvalidation & \textbf{100.00\%} & 17 & 10 & 14 & 17 \\
\rowcolor{toolbg}\propminimax & \textbf{100.00\%} & \textbf{24} & \textbf{17} & \textbf{20} & \textbf{25} \\
\bottomrule
\end{tabular}

\end{table}

\subsection{RQ4: Ablation Study}
\label{sec:rq4}
We finally examine how property skeletons, validation, and feedback each contribute to generating effective and comprehensive PBTs with semantic oracles, \ie\ valid tests that cover diverse operators, properties, and data generators.
Starting from \agentminimax, we add the three components one at a time, which gives \plusskeleton, \plusvalidation, and \propminimax, and we run the two variants on ONNX-MLIR for six hours.
Table~\ref{tab:rq4-ablation} reports the validity of their tests and the diversity of their valid tests, measured as in RQ1 and RQ2.
Skeletons contribute most of the gain in validity, 75.15 of the 81.82 percentage points between \agentminimax and \propminimax, and feedback contributes most of the gain in combinations.

\MyPara{Property Skeletons.}
Skeletons raise validity from 18.18\% to 93.33\%, and \plusskeleton uses 4.50$\times$ the properties and 10.00$\times$ the data generators of \agentminimax.
For example, a valid test of \plusskeleton checks that \codefont{Abs}, which returns the magnitude of each value, scales its output by the same positive constant as its input (\ie\ \codefont{Abs(3 * x) == 3 * Abs(x)}).
This relation holds for every input, and on an input that contains \codefont{-0.0} and \codefont{-3.75}, both sides of the compiled model return the expected 0.0 and 11.25.
This is because the skeleton of positive homogeneity requires a nonnegative constant, and the agent builds a small model around the selected operator, here two \codefont{Abs} and two \codefont{Mul} nodes, which passes the ONNX checker.
In contrast, 36.36\% of the tests of \agentminimax are invalid because they write a whole graph in which a \codefont{ReduceMean} node receives the axes to average over as a second input.
Their models declare opset 13, where \codefont{ReduceMean} takes these axes only as an attribute, so the ONNX checker rejects each model before compilation.

\MyPara{Validation.}
Validation raises validity from 93.33\% to 100.00\%, so every test of \plusvalidation is valid.
For example, one invalid test of \plusskeleton checks that two values that \codefont{Equal} reports as equal also have equal reciprocals (\ie\ \codefont{Equal(x, y)} implies \codefont{1/x == 1/y}).
IEEE~754 treats \codefont{+0.0} and \codefont{-0.0} as equal, but their reciprocals are $+\infty$ and $-\infty$, which are not equal, so this property does not apply to \codefont{Equal} and the test is invalid.
This is because \plusskeleton executes each candidate as soon as the agent generates it, while \plusvalidation first has a validation agent review whether the property applies to the selected operator, and skips each candidate that this review rejects, as described in Section~\ref{sec:validation}.

\MyPara{Feedback.}
Feedback raises the combinations of valid tests to 1.47$\times$ those of \plusvalidation, and \propminimax uses seven more properties.
For example, reference consistency, which compares compiled outputs with reference outputs, accounts for 29.41\% of the valid tests of \plusvalidation, while no property accounts for more than 12.00\% of the valid tests of \propminimax.
This is because usage records tell the agent how often each property has been tested, and prioritization rules steer it toward the untested ones, as described in Section~\ref{sec:execution-feedback}.
Feedback also keeps the agent from losing properties to repeated rejections.
\plusvalidation never learns why a candidate is rejected, so it generates consecutive candidates whose decision records (Section~\ref{sec:controlled-generation}) leave out the expected oracle, \ie\ the outcome that their assertions should observe, although their scripts contain these assertions.
The validator rejects all of them, including checks of associativity for \codefont{MatMul}, distributivity for \codefont{Mul}, and monotonicity for \codefont{Max}, and none of these three properties appears in the valid tests of \plusvalidation.
In contrast, \propminimax receives each rejection reason, so only 7.41\% of its candidates are rejected, against 41.38\% for \plusvalidation, and its valid tests use all three properties.
\begin{rqsummary}{4}
Each component of \tool contributes to generating more effective and comprehensive PBTs with semantic oracles.
Skeletons raise validity by 75.15 percentage points, validation raises it to 100.00\% by reviewing each candidate before execution, and feedback yields 1.47$\times$ the combinations by steering the agent toward untested properties and reducing rejected candidates.
\end{rqsummary}

\section{Related Work}
\label{sec:related-work}
\MyPara{Oracle Design for DL Compiler Testing.}
Prior work on testing DL compilers and libraries has made progress in generating valid and diverse test cases~\cite{nnsmith,neuri,gencog,deeprel,hirgen,modelmeta,synthfuzz,tzer,oatest}, and it judges these tests through crashes or differential comparison.
For example, Tzer~\cite{tzer} mutates TVM IR together with compiler pass sequences under coverage feedback, and it reports crashes, unexpected exceptions, and output differences between the unoptimized and optimized compilation of the same IR.
OATest~\cite{oatest} inserts optimization patterns from documented tests into seed graphs and applies differential testing to TVM and ONNX Runtime, and NNSmith~\cite{nnsmith} compares execution results across implementations.
\tool complements this work by guiding agents to construct executable semantic oracles from tensor algebra together with the computation graphs and tensor inputs they check.

\MyPara{Property-Based Testing and Metamorphic Testing.}
Property-based testing checks whether program behavior satisfies specified properties over generated inputs~\cite{quickcheck,uatpp}.
For example, UAT++~\cite{uatpp} augments model-based acceptance testing at WeChat Pay with random exploration and automatically synthesized properties.
Metamorphic testing addresses the oracle problem by checking relations between source and follow-up executions when exact expected outputs are hard to obtain~\cite{chen1998metamorphic,zhang2025llmmr,polyjuice,xiao2022mtdlcomp}.
LLMs can also help identify such relations.
For example, Zhang et al.~\cite{zhang2025llmmr} evaluate whether GPT-3.5 and GPT-4 can identify metamorphic relations for 37 software systems from prior metamorphic testing studies.
For tensor compilers, PolyJuice~\cite{polyjuice} uses equality saturation to generate equivalent computation graphs and compares their outputs to detect miscompilations.
MT-DLComp~\cite{xiao2022mtdlcomp} mutates real-world models by inserting subgraphs that always evaluate to zero, such as $\mathrm{ReLU}(-(i-j)^2)$ over the outputs $i$ and $j$ of two existing nodes, and checks that the compiled model and its mutant predict the same label with confidence scores within a tolerance.
These relations hold regardless of which operators a model contains, so they check whether compilation preserves a model's output, but not whether an individual operator computes the correct values.
\tool builds on this relational view and focuses on how an agent can instantiate a property across operators and compilers.
Each skeleton couples the relation with operator applicability conditions, tensor input constraints, and an oracle template.

\MyPara{Tensor and Linear Algebra Rules in DL Compilers.}
Algebraic identities from linear algebra have long been used to justify DL compiler optimizations and graph rewrites~\cite{taso,tensat,egg}.
For example, TASO~\cite{taso} automatically generates graph substitutions for DNN computation graphs from operator specifications, and TenSat~\cite{tensat} uses equality saturation to represent equivalent tensor graphs before extracting an optimized graph.
These approaches use algebraic rules to search for efficient implementations.
In \tool, the rules specify relationships that compiler executions should preserve.
For example, a reduction-decomposition oracle compares summing a concatenated tensor with summing its parts and adding the results.
The skeleton specifies the conditions on tensor shapes, reduction axes, and numerical comparisons, so the agent can construct a test for a target compiler under those conditions.

\MyPara{LLM- and Agent-Assisted Test Generation.}
Recent work has explored LLMs as test generators and fuzzing engines.
For example, TitanFuzz~\cite{titanfuzz} uses generative and infilling LLMs to generate and mutate valid DL programs for testing DL libraries.
FuzzGPT~\cite{fuzzgpt} further guides LLMs with historical bug-triggering programs to produce unusual edge cases.
Other work uses execution or coverage feedback to improve generated tests.
For example, CoverUp~\cite{coverup} prompts an LLM with coverage information, and TestForge~\cite{testforge} iteratively refines generated unit tests using execution and coverage feedback.
Gentoo~\cite{gentoo} uses a coding agent to iteratively write and refine JQF~\cite{JQF} input generators for Java fuzz targets, optionally with predicate-level coverage feedback.
Oracle-generation methods such as TOGLL~\cite{togll} also use LLMs to synthesize assertions.
However, an LLM may generate an assertion that treats a program's incorrect output as the expected result, so the test passes despite the bug~\cite{konstantinou2024llmsgeneratetestoracles}.

Recent work has also explored using LLMs to write property-based tests directly.
Vikram et al.~\cite{vikram2024largelanguagemodelswrite} report that LLMs can infer properties from API documentation, but their valid and sound PBTs cover only about a fifth of documented properties.
Agentic property-based testing~\cite{maaz2025agenticpropertybasedtesting} infers properties from docstrings, types, and documentation and analyzes test failures to refine tests and identify bugs in Python packages.
Naziri et al.~\cite{naziri2026multidomain} extend the evaluation of both approaches to ten programs in five domains, including DL libraries and the GCC and Clang compilers.
PBT-Bench~\cite{pbtbench} benchmarks coding agents on identifying properties from library documentation and writing property-based tests that expose semantic bugs injected into Python libraries.
For Android apps, PropGen~\cite{propgen} infers candidate functionalities from GUI states, synthesizes executable properties from their observed behavior, and revises imprecise properties based on execution feedback.
In \tool, reusable property skeletons supply the semantic relation and its applicability conditions before the agent generates code.
Controlled instantiation guides the agent to select compatible operators and data generators and implement the corresponding computation graphs, tensor inputs, and assertions.
Execution results, rejection reasons, and records of used combinations then guide subsequent instantiation.

Closest to our work, Lee et al.~\cite{lee2026propertytemplates} capture recurring property families of Apache Spark as property templates and combine Lean proofs over a model of PySpark with property-based testing.
\tool differs in domain, method, and properties: it targets DL compilers rather than data-intensive frameworks, validates instantiated tests instead of proving properties, and uses skeletons that share no properties with these templates, such as geometric decomposition along tensor axes and shift invariance along a normalization axis.
Moreover, whereas each Spark template fixes a test harness and its workload generators, the agent in \tool selects a data generator, specifies tensor shapes and value ranges, and adapts each oracle to the interface of each compiler.

\MyPara{Takeaways.}
\tool connects algebraic oracle specifications with agentic test generation through controlled instantiation.
The skeleton supplies reusable criteria for both generation and validation, while the agent implements the corresponding computation graphs, tensor inputs, and assertions for each compiler.
This shared semantic basis lets oracle knowledge carry across tests without requiring the agent to infer it anew for every operator.

\begin{table}[t]
\centering
\caption{Validity judgments of two independent LLM reviewers on the 361 cases that both review.
The reviewers agree on 358 cases (Cohen's $\kappa=0.94$).}
\label{tab:rq1-agreement}
\footnotesize
\setlength{\tabcolsep}{7pt}
\renewcommand{\arraystretch}{1.15}
\begin{tabular}{@{}lcc@{}}
\toprule
& \multicolumn{2}{c}{\textbf{GPT-6.1 Reviewer}} \\
\cmidrule(l){2-3}
\textbf{Opus-5.5 Reviewer} & \textbf{Valid} & \textbf{Invalid} \\
\midrule
\textbf{Valid} & 331 & 1 \\
\textbf{Invalid} & 2 & 27 \\
\bottomrule
\end{tabular}

\end{table}

\section{Threats to Validity}
\label{sec:threats}
\MyPara{Internal Validity.}
The reliability of the semantic oracles instantiated by \tool depends on the quality of its property skeletons.
To strengthen these specifications, we consulted researchers with expertise in tensor algebra and DL compiler testing when constructing the skeletons.
Generation and validation also use the same LLM, so both may overlook the same incorrect assumption about an operator or numerical comparison.
We therefore check each generated oracle on NumPy or ONNX's \codefont{ReferenceEvaluator} during evaluation.
Scripts decide the validity and alarm verdict of most tests, and LLM reviewers judge the tests that scripts cannot decide, such as every alarm and every test whose run result contradicts the reference.
Opus-5.5 reviews all of these tests, and GPT-6.1 independently reviews the cases that determine our counts or that the first review finds uncertain, such as one representative of each root cause, every false positive, and low-confidence cases.
When the two reviewers disagree, a third LLM, Fable-5.1, decides the case.
We use Cohen's kappa coefficient~\cite{cohen-kappa} to measure the agreement between the two reviewers.
As Table~\ref{tab:rq1-agreement} shows, they agree on the validity of 358 of 361 cases ($\kappa=0.94$) and on the alarm verdict of 330 ($\kappa=0.87$), which indicates that the choice of reviewer does not significantly affect our results.
The campaigns of the three open-weight agents go through the same review, where the two reviewers agree on the validity of 19 of 22 cases and on the alarm verdict of 20, and Fable-5.1 decides the three cases with a disagreement.
We then group compiler failures by root cause to avoid counting repeated manifestations as distinct bugs.

\MyPara{Construct Validity.}
Test validity and diversity do not fully capture the strength of generated semantic oracles.
For example, a valid test may check only output shapes without checking its selected algebraic property, so counting selected properties can overstate the diversity of implemented semantic checks.
We therefore inspect test assertions and assess issue replication and discovered compiler bugs alongside validity and diversity.
Reference limitations also affect validity estimates.
We conservatively classify tests with unverifiable oracles as invalid, which may underestimate the proportion of valid tests.
Whether a compiler failure on a legal input is a fault or an expected rejection can be ambiguous.
We count it as an expected rejection when the compiler explicitly declares the operator or form unsupported, and as a fault when the compiler accepts the operator in its own support checks and then fails to lower the given form.
We count a test that exposes numerical instability as valid when a correct algorithm stays within its oracle's tolerance in the declared data type, and we report these tests separately from the fault counts.
We count the operators that a test's property check targets, using \tool's target operator and, for \agentonly, the operators identified from its assertions, with all operators of a graph when an assertion checks the whole graph.

\MyPara{External Validity.}
Our evaluation covers TVM, ONNX-MLIR, and GeneSys, which differ in their input formats, compiler stacks, and backend assumptions.
Results may differ for other compiler versions, backends, or LLMs.
The six agents use four harnesses, namely Codex, Claude Code, Qwen Code, and DeepSeek Harness, so a difference between two agents reflects both the LLM and the harness.
We therefore compare \tool and \agentonly with the same agent, where both approaches share the harness, the tools, the reasoning effort, and the time budget.
Issue replication uses TVM cases, so its quantitative results may not generalize to other compilers or fault categories.
The current 16 property skeletons and 20 data generators also cover a limited set of semantic relations and tensor-input patterns.
This knowledge base can be readily extended by adding skeleton and generator documents.

\MyPara{Conclusion Validity.}
LLM randomness and prompt wording can affect which tests are generated within a fixed time budget.
We use the same agent for \tool and \agentonly and run each configuration for six hours per compiler.
Individual runs may differ in the operators explored and faults exposed, so we report trends that hold across six agents and three compilers and interpret the differences within the evaluated settings and time budgets.

\section{Conclusion}
\label{sec:conclusion}
Valid computation graphs and tensor inputs let tests reach a DL compiler's transformations, but they do not specify whether the compiled outputs are correct, and prior fuzzing work has largely judged these outputs through crashes or comparison with another implementation.
We presented \tool, an agentic testing framework that guides agents to construct semantic oracles from tensor algebra.
Its core insight is that the relation an oracle checks and the conditions that make it valid can be recorded once in a property skeleton and reused for every generated test.
Through controlled instantiation, the agent first selects an operator, a property, and a data generator, and then constructs the computation graph and tensor inputs on which the property must hold.
The skeleton guides each instantiation by recording the relation together with its operator applicability conditions, tensor input constraints, and an oracle template.
Validation checks each instantiation against its skeleton before execution, and feedback from validation, execution, and usage records steers subsequent generation toward untested combinations.

Compared with the same agent without \tool, controlled instantiation changes both what the agent tests and how reliably it tests.
Left to write tests freely, the agent concentrates on familiar operators and properties: 94.05\% of the valid TVM tests of \agentonly with GPT-6.1 target \codefont{add}, and across all six agents, six of the 17 property rows on TVM remain empty for \agentonly.
\tool covers at least 2.73$\times$ the operator--property pairs of \agentonly on each compiler and generates 85.34\% fewer invalid tests.
In our ablation with MiniMax-3.1 on ONNX-MLIR, skeletons alone raise validity from 18.18\% to 93.33\%.

Skeletons also change the oracle.
Every symptom that the tests of \agentonly expose is a crash, hang, or exception caught by the implicit oracle, and none is detected by an assertion that the agent wrote.
In contrast, the tests of \tool detect 96 distinct compiler root causes, 7.38$\times$ those of \agentonly, including 25 that produce silent wrong outputs, none of which \agentonly detects.
In issue replication, \tool detects 100 of 161 faulty TVM versions, against 53 for \agentonly with its outdated API calls repaired, and crash detection alone accounts for only five of the 100.
More broadly, \tool suggests that giving an agent reusable oracle knowledge, rather than leaving it to infer what to check, is what turns generated tests into semantic tests of a compiler.

\bibliographystyle{ACM-Reference-Format}
\bibliography{main}

\end{document}